\documentclass[12pt]{article}
\usepackage{eqsection,latexsym,epsf}
\usepackage{amsfonts}
\usepackage{amssymb}
\usepackage{epsfig}

\begin{document}

\title{Turbo codes: the phase transition.}

\author{
  \\
  {\small Andrea Montanari
\footnote{Address untill May 2000: 
Laboratoire de Physique Th{\'e}orique de l' Ecole Normale
Sup{\'e}rieure, {\it 24 rue Lhomond, 75231 Paris CEDEX 05, France.} }
}              \\[-0.2cm]
  {\small\it Scuola Normale Superiore and INFN -- Sezione di Pisa}  \\[-0.2cm]
 {\small\it I-56100 Pisa, ITALIA}          \\[-0.2cm]
  {\small Internet: {\tt  montanar@cibs.sns.it}}\\[-0.2cm]
  {\protect\makebox[5in]{\quad}}  
 \\
}
\vspace{0.5cm}

\def\tg{\mbox{\boldmath $\tau$}}
\def\sg{\mbox{\boldmath $\sigma$}}
\def\thg{\mbox{\boldmath $\theta$}}
\def\Gg{\mbox{\boldmath $\Gamma$}}
\def\Hg{\mbox{\boldmath $H$}}
\def\Jg{\mbox{\boldmath $J$}}
\def\Cg{\mbox{\boldmath $C$}}
\def\Bg{\mbox{\boldmath $B$}}
\def\xg{\mbox{\boldmath $x$}}
\def\yg{\mbox{\boldmath $y$}}
\def\xxg{\mbox{\boldmath $X$}}
\def\zb{\mbox{\boldmath $0$}}
\def\<{\langle}
\def\>{\rangle}
\def\arctanh{\mbox{arctanh}}

\maketitle
\thispagestyle{empty}   

\vspace{0.2cm}

\begin{abstract}
Turbo codes are a very efficient method for communicating reliably
through a noisy channel. There is no theoretical understanding
of their effectiveness. In Ref. \cite{Turbo1} they are mapped 
onto a class of disordered spin models. The 
analytical calculations concerning these models are reported here. 
We prove the existence of a no-error phase and compute its local 
stability threshold. As a byproduct, we gain some insight into the
dynamics of the decoding algorithm.
\end{abstract}

\vspace{4.2cm}
\begin{flushleft} 
LPTENS 00/13 
\end{flushleft}

\clearpage

\section{Introduction.}

Communication through a noisy channel is a central problem in Information 
Theory \cite{Cover}. Error correcting codes are a widespread method 
for compensating the information corruption due to the noise, by cleverly
increasing the redundancy of the message.
Turbo codes \cite{PrimoBerrou,Berrou2,Berrou3} are a recently 
invented class of error correcting codes with nearly optimal performances.
They allows reliable communication (i.e. very low error per bit probability)
with practical communication rates.

It is known, since the work of Sourlas
\cite{Sourlas1,Sourlas2,Sourlas3,Sourlas4}, that there exists a close 
relationship between the statistical behavior of error correcting codes and
the physics of some disordered spin models. Recently the
tools developed in statistical physics have been employed in 
studying Gallager-type codes \cite{Gallager, MacKay, Kanter}. 

In Ref. \cite{Turbo1} the equivalence discovered by Sourlas 
is extended to turbo codes, and the 
basic features of the corresponding spin models are 
outlined. A remarkable property of a large family of 
turbo codes, presented in Ref. \cite{Turbo1}, is the existence of a 
no-error phase. In other words the error probability per bit vanishes
beyond some critical (finite) signal to noise ratio. 
In Ref. \cite{Turbo1} some intuitive arguments supporting this thesis 
are given. Some analytical results concerning the critical value of
the signal to noise ratio are announced without giving any derivation.
These results are compared with numerical simulations.

In this paper we present the analytical results in their full generality,
and explain their derivation. We prove the existence of the no-error
phase and find the condition for its local stability. 
This condition is derived in two different approaches. In the first
one we study the asymptotic dynamics of the decoding algorithm.
In the second approach we use replicas and establish the condition
for stability in the full replica space.
Local stability is a necessary but not sufficient condition for the
stability of the no-error phase. The critical signal to noise ratio obtained 
from local stability is the correct one only if the phase transition is 
a second order one: in the general case it is only a lower bound. 

The spin models which are equivalent to turbo codes
have the following statistical weight \cite{Turbo1}:
\begin{eqnarray}
{\cal P}(\sg^{(1)},\sg^{(2)}|\Jg,\beta) & \equiv & \frac{1}{Z(\Jg,\beta)}
\prod_{i=1}^N\delta\left(\epsilon_{\rho(i)}(\sg^{(1)}),
\epsilon_{i}(\sg^{(2)})\right)
e^{-\beta \sum_{k=1}^2 H^{(k)}(\sg^{(k)})}\label{Models1}\\
H^{(k)}(\sg)& \equiv & -\sum_{i=1}^N J^{(k)}_i \epsilon_{i}(\sg)
-\sum_{i=1}^N h^{(k)}_i \eta_{i}(\sg)
\label{Models2}
\end{eqnarray}
The dynamical variables of the model are the spins 
$\sg^{(k)}\equiv\{\sigma^{(k)}_1,\dots,\sigma^{(k)}_N\}$ with
$k=1,2$. We shall choose them to be Ising spins
\footnote{This corresponds to considering codes which works with a
binary alphabet.}, 
that is $\sigma^{(k)}_i = \pm 1$. 
The spins enters in the hamiltonians $H^{(k)}(\sg)$ through the
local interaction terms $\epsilon_{i}(\sg)$ and $\eta_{i}(\sg)$ which
are products of $\sigma$'s.
Their exact form can be encoded in two set of numbers 
$\kappa(j;1) = 0,1$ and $\kappa(j;2) = 0,1$ as follows:
$\epsilon_i(\sg)\equiv \prod_{j=0}^r\sigma_{i-j}^{\kappa(j;1)}$ and 
$\eta_i(\sg)\equiv \prod_{j=0}^r\sigma_{i-j}^{\kappa(j;2)}$.
In order to fix completely our notation we set $\kappa(0;1)=\kappa(0;2)=1$.
The quenched variables are: 
\begin{itemize}
\item the couplings
$\Jg\equiv\{J^{(k)}_i;h^{(k)}_i\}$, whose distribution 
${\cal P}(\Jg)\equiv\prod_{i,k} P(J^{(k)}_i)\;P(h^{(k)}_i)$ 
satisfies the conditions
$\int\!dJ^{(k)}_i\;P(J^{(k)}_i)\;J^{(k)}_i>0$ and 
$\int\!dh^{(k)}_i\;P(h^{(k)}_i)\;h^{(k)}_i>0$; 
\item the permutation $\rho:\{1,\dots,N\}\to\{1,\dots,N\}$, which
has uniform distribution.
\end{itemize}
It is convenient to impose a fixed boundary condition
at one end of the chain  (i.e. $\sigma_i = +1$ for $i\le 0$) and a free
boundary condition at the other end. 
The model is composed by two one dimensional substructures (chains),
which interact
through the Kronecker delta functions in Eq. (\ref{Models1}). When 
the average over permutations is taken into account this interaction turns
into a mean field one. This interplay between the two subsystems, 
each one possessing a one dimensional structure, 
and the mean field interaction which couples
them is clearly displayed by the analytical calculations.
For further explanations on Eqs. (\ref{Models1}-\ref{Models2}) and their
motivation we refer to \cite{Turbo1}.

The paper is organised as follows. In Sections 
\ref{WithoutReplicas1} and \ref{WithoutReplicas2}
we present a first derivation of the stability condition. We write a ``mean
field'' equation which describes the dynamics of the decoding
algorithm (Sec. \ref{WithoutReplicas1}),
we show that it possesses a no-error fixed point and then study its 
behavior in a neighbourhood of this fixed point (Sec. \ref{WithoutReplicas2}).
Thanks to this derivation we will understand how this fixed point is reached.
In Section \ref{WithReplicas} replicas are introduced in order to 
compute the average over the permutations. We exhibit the 
no-error saddle point. In Section \ref{Stability} the stability
of the no-error saddle point is studied by diagonalizing the second derivative
of the free energy. Finally in Section \ref{Conclusion} the validity of
our calculations is discussed.
Appendix \ref{AppendiceAlgebra} collects some useful (although simple) facts
of algebra. In Appendix \ref{AppendiceProbabilita} the type of integral 
equations which appear in Section \ref{WithoutReplicas2} is studied in detail.
%
%
\section{A ``mean field'' equation for the decoding algorithm.}

\label{WithoutReplicas1}

\newtheorem{lemmicitati}{Lemma}[section]

Some properties concerning the models defined by Eqs.
(\ref{Models1}-\ref{Models2}) can be obtained by 
considering the ``turbo decoding'' algorithm and making some factorization 
hypothesis. These hypothesis enable us to obtain a recursive 
integral equation for the probability distribution of
a local field. They can be justified on heuristic grounds 
and arguments of this kind will be given later in this Section.
Moreover the replica calculation presented in the Section
\ref{WithReplicas} does support our arguments.
In particular this approach allows us to derive the critical noise
below which ``perfect'' decoding is possible.

Turbo decoding is an iterative algorithm. 
The iteration variables are the fields 
$\Gg^{(k)} \equiv \{\Gamma^{(k)}_1,\dots,\Gamma^{(k)}_N\}$ with $k=1,2$.
The step $t$ of the turbo decoding algorithm is defined as 
follows \cite{Turbo1}:
\begin{eqnarray}
\Gamma_i^{(1)}(t+1) & = & \frac{1}{\beta}\arctanh\left[
\<\epsilon_{\rho^{-1}(i)}(\sg)\>^{(2)}_{\Gamma^{(2)}(t)}\right]
-\Gamma_{\rho^{-1}(i)}^{(2)}(t)\label{TurboDecoding1}\\
\Gamma_i^{(2)}(t+1) & = & \frac{1}{\beta}\arctanh\left[
\<\epsilon_{\rho(i)}(\sg)\>^{(1)}_{\Gamma^{(1)}(t)}\right]
-\Gamma_{\rho(i)}^{(1)}(t)\label{TurboDecoding2}
\end{eqnarray}
The expectation value $\<\cdot \>^{(k)}_{\Gamma^{(k)}}$ is intended to be taken
with respect to the Boltzmann weight with the modified hamiltonian
$H^{(k)}(\sg)-\sum_{i=1}^N \Gamma_i^{(k)}\epsilon_i(\sg)$.
The iteration variables $\Gamma^{(k)}_i$ should be interpreted as external 
fields conjugate to the operators $\epsilon_i(\sg^{(k)})$. They describe, in 
an approximate way, the action of each of the two chains on the other one.

In order to lighten the notation, let us write Eqs.
(\ref{TurboDecoding1}-\ref{TurboDecoding2}) in the form:
\begin{eqnarray}
\Gg ^{(k)}(t+1) = F^{(k)}_{\rho}\left(\Gg ^{(k')}(t),\Jg^{(k')}\right)
\label{TurboDecodingShort}
\end{eqnarray}
with $k' = 2$ if $k = 1$, and $k' = 1$ if $k = 2$. 
Due to the randomness in the couplings $\Jg$, the fields $\Gg$ are random 
variables. 
Equation (\ref{TurboDecodingShort}) implies an integral equation for 
the probability distribution of $\Gg$:
\begin{eqnarray}
{\cal P}_{t+1}(\Gg^{(k)}) = \int d\Gg^{(k')}\int d\Jg^{(k')}
{\cal P}_t(\Gg^{(k')},\Jg^{(k')})
\delta\left[\Gg ^{(k)} - 
F^{(k)}_{\rho}\left(\Gg ^{(k')},\Jg^{(k')}\right)\right]
\label{ProbabiltyIteration}
\end{eqnarray}
Let us state a few approximations which allow us to reduce Eq. 
(\ref{ProbabiltyIteration}) to a much simpler one. 
\begin{itemize}
\item[\bf{(1)}] We make the substitution 
${\cal P}_t(\Gg^{(k')},\Jg^{(k')})\to 
{\cal P}_t(\Gg^{(k')}){\cal P}(\Jg^{(k')})$ in Eq. (\ref{ProbabiltyIteration}).
This yields a closed integral equation 
describing the evolution of the distribution ${\cal P}_t(\Gg^{(k)})$.
\item[\bf{(2)}]  We neglect correlations between the fields at different sites:
\begin{eqnarray}
{\cal P}_t(\Gg^{(k)})\simeq\prod_{i=1}^N \pi^{(k)}_{i,t}(\Gamma^{(k)}_i)
\label{HypothesisTwo}
\end{eqnarray}
\end{itemize}
These two hypothesis imply that Eq. (\ref{ProbabiltyIteration}) is equivalent 
to:
\begin{eqnarray}
\pi^{(k)}_{i,t+1}(y) &= &\int_{-\infty}^{+\infty}\!\!
d\pi^{(k')}_t[\xg ]\int\!d{\cal P}[\Jg]\;\delta\left(y-
\frac{1}{\beta}\arctanh\left(\<\epsilon_{\hat{\rho}(i)}(\sg)\>_{\Jg,\xg}
\right)+x_{\hat{\rho}(i)}\right)
\label{Pdependent}\\
d\pi^{(k')}_t[\xg ]&\equiv&\prod_{i=1}^N dx_i \;\pi^{(k')}_t(x_i)
\end{eqnarray}
where $\hat{\rho}$ is the appropriate permutation of $\{1,\dots,N\}$, 
i.e. $\hat{\rho}=\rho^{-1}$ if $k=1$ and $\hat{\rho}=\rho$ if
$k=2$. The expectation value 
$\<\cdot \>_{\Jg,\xg}$ on the right hand side of 
Eq. (\ref{Pdependent}) 
has to be taken with respect to the hamiltonian 
$H(\sg) \equiv -\sum_{i=1}^N (J_i+x_i) \epsilon_{i}(\sg)
-\sum_{i=1}^N h_i \eta_{i}(\sg)$.

Let us now define a field distribution averaged over the permutations and the 
sites:
\begin{eqnarray}
\pi^{(k)}_t(x)\equiv \frac{1}{N!}\sum_{\rho}\frac{1}{N}\sum_{i=1}^N 
\pi^{(k)}_{i,t}(x|\rho)
\end{eqnarray}
where we made explicit the dependence of $\pi^{(k)}_{i,t}$ upon the specific 
permutation $\rho$ which defines the code. 
We can now state our last approximation.
\begin{itemize}
\item[\bf{(3)}] We make the substitution 
$\pi^{(k)}_{i,t}(x|\rho)\to \pi^{(k)}_t(x)$ on the right hand side 
of Eq. (\ref{Pdependent}).
\end{itemize}
This yields a recursive equation for $\pi^{(k)}_t$:
\begin{eqnarray}
\pi_{t+1}(y) = \frac{1}{N}\sum_{i=1}^N\int_{-\infty}^{+\infty} 
d\pi_t[\xg ]\int d{\cal P}[\Jg]\ \delta\left(y-
\frac{1}{\beta}\arctanh\left(\<\epsilon_{i}(\sg)\>_{\Jg,\xg}\right)
+x_{i}\right)
\label{Pindependent}
\end{eqnarray}
The indices $(k)$ and $(k')$ have been dropped since we can define 
$\pi_t=\pi^{(1)}_t$ for $t$ odd, and $\pi_t=\pi^{(2)}_t$ for $t$ even, or 
vice-versa.
A byproduct of this heuristic derivation is the expression
for the probability distribution of the expectation values
$\<\epsilon_{i}(\sg)\>$ after $t$ iterations of the turbo decoding
algorithm: ${\cal P}_t(\epsilon) = \frac{1}{N}\sum_{i=1}^N
\int_{-\infty}^{+\infty} 
d\pi_t[\xg ]\int d{\cal P}[\Jg]\ \delta\left(\epsilon-
\<\epsilon_{i}(\sg)\>_{\Jg,\xg}\right)$.

Let us discuss the validity of the approximations made in deriving
Eq. (\ref{Pindependent}).
\begin{itemize}
\item[{\bf (1)} and {\bf (2)}] These approximations 
should be accurate in the thermodynamic limit for a generic random
permutation $\rho$.
The reason is that the correlations produced by Eqs. 
(\ref{TurboDecoding1}-\ref{TurboDecoding2}) have short range: 
$\<\epsilon_i(\sg)\>$ and $\<\epsilon_j(\sg)\>$
have a significant correlation only if $|i-j|$ is less than some 
characteristic length. 
The random permutation $\rho$ reshuffles the sites so that
the correlation between two fields $\Gamma^{(k)}_i$ and $\Gamma^{(k)}_j$
is vanishing with high probability if $|i-j|$ is required to be ``small''.
The correlations which ``survive'' (non vanishing only between 
``distant'' sites) are irrelevant when computing the expectation 
values of local operators. In order to make this last assertion plausible, let 
us suppose that, for each site $i$, we can find a ``large'' 
\footnote{Here ``large'' means that $\lim_{N\to\infty}L(N)=\infty$.}
interval $[i-L(N),i+L(N)]$ of the chain, such that the correlations
between the couplings inside the interval are negligible.
The expectation value $\<\epsilon_{i}(\sg)\>_{\Jg,\xg}$ will not depend
(as $N\to\infty$) upon the couplings outside $[i-L(N),i+L(N)]$  
(this is always true in one dimension at non zero temperature) 
and can be then safely computed 
without taking into account the correlations.
It is easy to find a similar argument concerning the correlations between 
$\Gg^{(k')}$ and $\Jg^{(k')}$ in Eq. (\ref{ProbabiltyIteration}). 
\item[\bf{(3)}] This is the probabilistic analogue of the 
replica symmetric approximation. 
Let us consider the fixed point equation $\pi_{t+1}=\pi_t$
corresponding to the dynamics defined by Eq. (\ref{Pindependent}). 
It is remarkable that this fixed point equation
coincides with the saddle point equation obtained by the
standard replica method in the replica symmetric approximation 
(see Section \ref{WithReplicas}).
This fact confirms our conclusions about the relevance of the various 
approximations.
\end{itemize}
%
%
\section{The behavior of the decoding algorithm.}

\label{WithoutReplicas2}

Equation (\ref{Pindependent}) is the final outcome of our heuristic
derivation.  We want to study its behavior 
when the distribution $\pi(x)$ is concentrated on large 
values of the field $x$, that is when the error probability is very
small.
In this regime the most relevant spin configuration satisfies 
$\epsilon_i(\sg)=+1$  for each $i=1,\dots,N$.
The lowest excitations are such that $\epsilon_i(\sg)=-1$
only on a few sites. 
The first crucial point will be to understand that, for 
a class of hamiltonians of the type (\ref{Models2}) 
(which will be defined as ``recursive ''), the 
energy to be paid for flipping a single $\epsilon$ variable
diverges in the thermodynamic limit.
The second point will be to evaluate the energy to be paid for
flipping two $\epsilon$ variables. In order to treat both these passages
in full generality it is convenient to use an algebraic bookkeeping technique
which we shall soon explain. The results concerning these two
points will be useful again in Section \ref{Stability}.

A preliminary step consists in making the change of variables 
$X_i\equiv e^{-2\beta x_i}$ and introducing the corresponding distribution 
function $Q_t(X)dX = \pi_t(x)dx$. Low $X$'s correspond then to 
large local fields, i.e. to low error probability.
The result is
\begin{eqnarray}
Q_{t+1}(Y) = \frac{1}{N}\sum_{i=1}^N\int_0^{\infty} 
dQ_t[\xxg ]\int d{\cal P}[\Jg]\ \delta\left(Y-
\frac{1}{X_i}\,\frac{Z(\epsilon_i(\sg)=-1;\Jg,\xxg)}
{Z(\epsilon_i(\sg)=+1;\Jg,\xxg)}\right)
\label{Riscritta}
\end{eqnarray}
where
\begin{eqnarray}
Z(\epsilon_i(\sg)=\epsilon;\Jg,\xxg)\equiv Z_i(\epsilon)=
\sum_{\sg:\epsilon_i(\sg)=\epsilon}
e^{-\beta H(\sg)}
\prod_{k=1}^NX_k^{\frac{1}{2}(1-\epsilon_k(\sg))}
\end{eqnarray}
with $H(\sg) = -\sum_i J_i \epsilon_{i}(\sg)-\sum_i h_i \eta_{i}(\sg)$.
Let us introduce some notations in order to write down the small $X$ 
expansion of $Z_i(\epsilon)$: $(k_1,\dots,k_l)$ is an $l$-uple (not ordered)
of integers in $\{1,\dots,i-1,i+1,\dots,N\}$; $\sg_0$ is the configuration
such that $\epsilon_i(\sg)=+1$ for all the sites $i$;  
$\sg(k,l,m,\dots)$ is the configuration such that $\epsilon_j(\sg) = -1$ 
if $j=k,l,m,\dots$ and $\epsilon_j(\sg) = 1$ otherwise 
(there is at most one such configuration once 
the boundary conditions have been specified); $E_0 \equiv H(\sg_0)$ is the 
energy of the ordered configuration; finally
$\Delta(k,l,m,\dots) \equiv H(\sg(k,l,m,\dots))-H(\sg_0)$. 
The following expressions are 
straightforward: 
\begin{eqnarray}
Z_i(+1) = e^{-\beta E_0}\sum_{l=0}^{N-1}\sum_{(k_1,\dots,k_l)}
X_{k_1}\dots X_{k_l}e^{-\beta \Delta(k_1,\dots,k_l)}
\label{LowX1}\\
Z_i(-1) = X_i e^{-\beta E_0} \sum_{l=0}^{N-1}\sum_{(k_1,\dots,k_l)}
X_{k_1}\dots X_{k_l}e^{-\beta \Delta(i,k_1,\dots,k_l)}
\label{LowX2}
\end{eqnarray}

The ``bookkeeping technique '' which we shall adopt
in treating the above expansions consists in using the
algebra of ``generating polynomials'' \cite{Turbo1}.
This approach allows us to consider a general hamiltonian of the type 
(\ref{Models2}). 
Let us define the following polynomials on $\mathbb{Z}_2$: 
$G(x) \equiv \sum_{j=1}^{\infty} G_j x^j$, with $\sigma_j=(-1)^{G_j}$;  
$g_n(x) = \sum_{j=0}^r \kappa(j;n)x^j$; 
${\cal G}^{(n)}(x) \equiv g_n(x)\cdot G(x)\equiv 
\sum_{j=1}^{\infty} {\cal G}^{(n)}_j x^j$. Notice that the boundary condition 
on $\sg$ can be translated as follows: $G(x)$ is a series of strictly positive
powers of $x$.

It is necessary to distinguish two types of models:
in the first case $g_1(x)$ divides $g_2(x)$, i.e. $g_2(x)/g_1(x)$ is a 
polynomial (these are the ``non recursive''
models, a particular case being $\epsilon_i(\sg) = \sigma_i$); 
in the second one $g_1(x)$ does not divide $g_2(x)$, i.e. 
$g_2(x)/g_1(x)$ is a series (``recursive'' models). 

We shall treat the ``recursive'' models first.
In this case the first order terms in the expansions (\ref{LowX1}) and 
(\ref{LowX2}) are exponentially small in the size. 
In order to prove this assertion, let us consider the configuration $\sg(l)$. 
The relevant generating polynomials are ${\cal G}^{(1)}(x) = x^l$ and  
${\cal G}^{(2)}(x) = x^l g_2(x)/g_1(x)$. The form of ${\cal G}^{(2)}(x)$
is given by the following result of algebra
\begin{lemmicitati}
\label{Eccitazioni1}
Let $g(x)$ and $f(x)$ be two polynomials on $\mathbb{Z}_2$ such that 
$g(0) = f(0) =1$, $f(x)\not\equiv 1$, and 
their greatest common divisor $\gcd(f(x),g(x))$ is equal to 1. 
Then there exists an integer $\omega$ such that 
$g(x)/f(x)=\sum_{n=0}^{\infty}x^{n\omega}p_n(x)$ with 
$\deg[p_n(x)]<\omega$ and
$p_n(x) = p_{\infty}(x)\neq 0$ if $n$ is large enough. 
Hereafter we shall call $\omega(f)$ the smallest of such integers.
\end{lemmicitati} 
An explicit expression for $\omega(f)$ is given in the Appendix
\ref{AppendiceAlgebra}. The Lemma
\ref{Eccitazioni1} applies to our case if we divide both $g_1(x)$
and $g_2(x)$ by their greater common divisor: 
$f_k(x)\equiv g_k(x)/\gcd(g_1(x),g_2(x))$, so that $\gcd(f_1(x),f_2(x))=1$.
It implies that if we write down the numbers 
$\eta_j(\sg(i))=\pm 1$ we get an antiperiod followed by a non trivial 
periodic sequence with period $\omega(f_1)$. 
Let us consider a site ``in the bulk'': $N\delta <i<N(1-\delta)$ with 
$\delta$ a (small) positive number. Then, using the convention $h_j = 0$ 
for $j>N$, we get: 
\begin{eqnarray}
\Delta(i) & = & 2J_i+2\sum_{j=1}^N {\cal G}^{(2)}_j h_j = 
2J_i+2\sum_{n=0}^{\infty}\sum_{k=0}^{\omega(f_1)-1}p_{n,k}
h_{i+n\omega(f_1)+k}\label{UnFlip}
\end{eqnarray}
which diverges almost surely in the thermodynamic limit if $\<h\> >0$ 
(see the Introduction on this point).
In Eq. (\ref{Riscritta}) we must sum also terms which are ``near'' the
boundaries, i.e. $i\le N\delta$ or $i\ge N(1-\delta)$. These give however 
a negligible contribution.

Let us now consider the second order terms of the expansions (\ref{LowX1}) and 
(\ref{LowX2}).
They involve configurations $\sg(k,l)$ with two flipped
$\epsilon(\sg)$'s. The only
configurations which give a non negligible contribution are the ones which
involve a finite (in the $N\to\infty$ limit) number of flipped $\eta(\sg)$'s.
This corresponds to choosing $k$ and $l$ such that $(x^k+x^l)g_2(x)/g_1(x)$
is a polynomial (and not an infinite series).
The following useful result is proved in the Appendix \ref{AppendiceAlgebra}.
\begin{lemmicitati}
\label{Eccitazioni2possibili}
Let $f(x)$ be a polynomial on $\mathbb{Z}_2$ such that $f(0)=1$ and $k$ 
an integer. Then there exists an integer $\omega(f)$ such that 
$f(x)$ divides $1+x^k$ if and only if $k$ is a strictly positive multiple of
$\omega(f)$.
\end{lemmicitati} 
As suggested by the notation the $\omega(f)$'s cited in Lemmas
\ref{Eccitazioni1} and \ref{Eccitazioni2possibili} are indeed equal.
The terms which give a non vanishing contribution
at order $X^2$ in the expansions (\ref{LowX1}-\ref{LowX2}) are the ones 
corresponding to configurations $\sg(k,l)$ such that $|k-l|$ is a multiple 
of $\omega(f_1)$. In order to evaluate these terms we must count the number
of flipped $\eta(\sg)$'s. This number is nothing but the
number of non zero coefficients in the polynomial 
$(x^k+x^l)g_2(x)/g_1(x)$. Let us define the weight of a polynomial 
$p(x)=\sum_k p_kx^k$ over $\mathbb{Z}_2$ as the number of its
non zero coefficients: $\mbox{weight}(p) \equiv \#\;\{p_k|p_k\neq 0\}$.
The weight of $(x^k+x^l)g_2(x)/g_1(x)$ is given, for a large class of
hamiltonians, by the following lemma.
\begin{lemmicitati}
\label{Eccitazioni2energia}
Let $f(x)$ and $g(x)$ be two polynomial on $\mathbb{Z}_2$ such that 
$f(0) = g(0) = 1$, $f(x)\not\equiv 1$, and 
$\gcd(f(x),g(x)) = 1$. If $\deg(g)\le\omega(f)$ then
the weight of $s_m(x)\equiv (1+x^{m\omega(f)})g(x)/f(x)$ is given by
${\rm weight}(s_m) = w_0(f,g)+w_1(f,g) m$ for each $m\ge 1$.
The coefficients $w_0(f,g)$ and $w_1(f,g)$ are positive 
integers whose explicit expressions are given 
by Eqs. (\ref{Peso1}-\ref{Peso2}).
\end{lemmicitati}
Appendix \ref{AppendiceAlgebra} contains also
an illustration of what could happen in the more general case.

By using Lemmas \ref{Eccitazioni2possibili} and \ref{Eccitazioni2energia}
we can linearize with respect to $X$ the expression on the r.h.s. 
of Eq. (\ref{Riscritta}):
\begin{eqnarray}
\frac{1}{X_i}\,\frac{Z_i(-1)}{Z_i(+1)}  =  
\sum_{m\neq 0 }
X_{i+m\omega(f_1)}e^{-\beta\Delta(i,i+m\omega(f_1))}+O(X^2)
\label{Sviluppo}
\end{eqnarray}
and defining
$s_m(x) \equiv (1+x^{m\omega(f_1)})g_2(x)/g_1(x)=\sum_j s_{m,j} x^j$ we get
\begin{eqnarray}
\Delta(k,l)  = 
2J_k+2J_l+
2\sum_j s_{m,j} h_{\min(k,l)+j}
\end{eqnarray}
if $|k-l|=m\omega(f_1)$. Clearly Eq. (\ref{Sviluppo}) holds only
for $i$ in the ``bulk'' (i.e. $N\delta<i<(1-\delta)N$) up to terms which 
are exponentially small in the size $N$.

Our first important observation is that the right hand side of 
Eq. (\ref{Sviluppo}) vanishes if $X_k=0$
for $k=1,\dots,N$. This means that $Q_*(X) = \delta(X)$ is a fixed point of 
Eq. (\ref{Riscritta}) for ``recursive'' models. Recall that the change of 
variables which yields Eq. (\ref{Riscritta}) is $X=e^{-2\beta x}$ and that 
$x$ has the meaning of an effective field acting on $\epsilon_i(\sg)$. 
The solution $Q_*(X)$ corresponds then to a phase with completely frozen spins:
$\<\epsilon_i(\sg)\> = +1$.

We would like to understand if this phase is stable for some
temperature $\beta$ and some distribution of the couplings. 
A possible approach is to study the turbo decoding dynamics
(as described by Eq. (\ref{Riscritta})) when starting from a distribution
``near'' $Q_*(X)$. Let us suppose that, for $Q_t(X)$ near enough to $Q_*(X)$,
we can safely neglect $O(X^2)$ terms on the r.h.s. of Eq. (\ref{Sviluppo}):
\begin{eqnarray}
Q_{t+1}(Y) = \frac{1}{N}\sum_{i=1}^N\int_0^{\infty}\!
dQ_t[\xxg ]\int\! d{\cal P}[\Jg]\ \delta\left(Y-
\sum_{ m\neq 0 }
X_{i+m\omega(f_1)}e^{-\beta\Delta(i,i+m\omega(f_1))}\right)
\label{Linearizzata}
\end{eqnarray}
This equation is very similar to 
a class of recursive equations which appear in a completely different context:
polymers on disordered trees \cite{DerridaSpohn, BuffetTrees, DerridaRig,
DerridaCook1, DerridaCook2}. These are of the type
\begin{eqnarray}
P_{t+1}(Z) =\int_0^{\infty}\prod_{i=1}^K dZ_i\;P_t(Z_i)
\int \!\!\rho(V_1,\dots,V_K)\;dV_1\dots dV_K\;
\delta\left(Z-\sum_{i=1}^K e^{-\beta V_i}Z_i\right)
\label{Alberi}
\end{eqnarray}
The only non trivial difference is that the linear function of $\xxg$ 
appearing inside the delta function on the r.h.s. of Eq. (\ref{Linearizzata})
depends upon a macroscopic (indeed linear in $N$) number of $X$'s. 
In Eq. (\ref{Alberi}), instead, 
only a finite number of variables appears: $K$ is the
coordination number of the tree minus one. Notice however that,
for $m$ large, $\Delta(i,i+m\omega(f_1))\sim 2\;\mbox{weight}(s_m)\<h\>
\sim 2 w_1(f_1,f_2) m \<h\>$. We can thus truncate the sum in 
Eq. (\ref{Linearizzata}) to $m\le M$ by making an error of order 
$O(e^{-cM})$ and we guess that the limit $M\to\infty$
can be taken at the end without problems
\footnote{This argument is not mathematically
rigorous since it is not honest to use the central limit theorem in this case:
we refer to Appendix \ref{AppendiceProbabilita} for more convincing 
arguments.}.

Let us summarize some results of \cite{DerridaSpohn} which are useful in our
discussion. It turns out that Eq. (\ref{Alberi}) is equivalent to a 
discretization of the Kolmogorov-Petrovsky-Piscounov (KPP) equation 
\cite{KPP} (a well 
studied partial differential equation). Using this equivalence the large time 
limit of Eq. (\ref{Alberi}) is obtained:
\begin{eqnarray}
P_t(X)\rightarrow e^{-\beta c(\beta)t}\,\overline{P}(Xe^{-\beta c(\beta)t})
\label{Larget}
\end{eqnarray}
corresponding to a front wave solution of the KPP equation with front velocity 
$c(\beta)$. If we define the function
\begin{eqnarray}
v(\beta) \equiv 
\frac{1}{\beta}\log\left(\sum_{i=1}^K
\int \!\!dV_1\dots dV_K\;\rho(V_1,\dots,V_K)\;e^{-\beta V_i}
\right)
\end{eqnarray}
then the front velocity is given by the following construction:
\begin{eqnarray}
c(\beta) = 
\left\{
\begin{array}{ccc}
v(\beta) & \mbox{if} & \beta\le\beta_c\\
v(\beta_c) & \mbox{if} & \beta>\beta_c
\end{array}
\label{FrontVelocity}
\right.
\end{eqnarray}
with $\beta_c$ given by 
\begin{eqnarray}
\left.\frac{d}{d\beta}\right|_{\beta_c}v(\beta) =0
\label{CriticalTemperature}
\end{eqnarray}
At the critical temperature $\beta_c$ a freezing phenomenon takes place with
the front velocity sticking to its minimal value. 

Let us apply these results to our case, i.e. to Eq. (\ref{Linearizzata}). 
The large time solution 
$Q_t(X)\sim e^{-\beta c(\beta)t}\,\overline{Q}(Xe^{-\beta c(\beta)t})$  
gives the correct behavior for $t\to\infty$ only if $c(\beta)<0$. In this case 
$\lim_{t\to\infty} Q_t(X)= Q_*(X)$ and 
it is then correct to linearize Eq. (\ref{Riscritta}): the 
frozen phase is stable. If, on the other hand, $c(\beta)\ge 0$ then we must  
take into account higher order terms in the low $X$ expansion and 
the asymptotic form is no longer of the type defined by
Eq. (\ref{Larget}): the frozen phase is unstable.

In the thermodynamic limit we get 
\begin{eqnarray}
e^{\beta v(\beta)} & = &\sum_{m\neq 0}\
\int d{\cal P}[\Jg] e^{-\beta\Delta(i,i+m\omega(f_1))}=
\label{PreVelocity1}\\
& = & 2\left(\int\! dJ \;P(J)\; e^{-2\beta J}\right)^2
\sum_{m=1}^{\infty}\left(\int\! dh\; P(h)\; e^{-2\beta h}\right)
^{{\rm weight}(s_m)}
\nonumber
\end{eqnarray}
The front velocity $c(\beta)$ is obtained by applying the construction 
given in Eqs. (\ref{FrontVelocity}-\ref{CriticalTemperature}) to 
Eq. (\ref{PreVelocity1}).
If the hypothesis of Lemma \ref{Eccitazioni2energia} are satisfied
we can easily sum the series:
\begin{eqnarray}
e^{\beta v(\beta)} & = \frac{
\displaystyle 2\left(\int\! dJ \;P(J)\; e^{-2\beta J}\right)^2
\left(\int\! dh\; P(h)\; e^{-2\beta h}\right)^{w_0(f_1,f_2)+w_1(f_1,f_2)}}
{\displaystyle 
1-\left(\int\! dh\; P(h)\; e^{-2\beta h}\right)^{w_1(f_1,f_2)}}
\label{PreVelocity2}
\end{eqnarray}
We discuss now Eq. (\ref{PreVelocity2}), the more general case being completely
analogous.
The series converges only if $\int\! dh\; P(h)\; e^{-2\beta h}<1$. If
$\int\! dh\; P(h)\; h>0$, as we supposed sinc the beginning, 
then convergence is assured for $0<\beta<\beta_1$ with  
$\int\! dh\; P(h)\; e^{-2\beta_1 h}=1$. It is easy to see that
$\beta v(\beta)$ is strictly convex for $0<\beta<\beta_1$ and thus
$v(\beta)$ has either one global minimum or is strictly monotonic
for $0<\beta<\beta_1$. Since
$\lim_{\beta\to 0^+} v(\beta) = \lim_{\beta\to\beta_1^-} v(\beta) = +\infty$ 
the first possibility is excluded and we conclude that $0<\beta_c<\beta_1$.
The important point is that the right hand side of Eq. (\ref{PreVelocity1})
is well defined every time we need of it, i.e. for $0<\beta<\beta_c$.

In applications to turbo codes a simplification occurs: we are 
interested in a particular temperature, $\beta=1$, and we are left with a
unique parameter: the signal to noise ratio $1/w^2$. Moreover the probability 
distributions of the couplings are fixed by the characteristics of the
communication channel \cite{Sourlas1,Turbo1}. If we introduce the auxiliary 
variables $\hat{J}$ and $\hat{h}$, which correspond to the
output of the channel, the probability distributions are obtained as follows
\begin{eqnarray}
P(J)\;dJ = P(\hat{J}|+1)\;d\hat{J}\quad\mbox{with}\quad
J=\frac{1}{2}\log\frac{P(\hat{J}|+1)}{P(\hat{J}|-1)}
\end{eqnarray} 
where $P(\hat{J}|\tau)$ is the probability distribution of the output of 
the channel conditional to the input $\tau$. A similar expression holds for
$h$. If the channel is symmetric (i.e. if $P(\hat{J}|-1)=P(-\hat{J}|+1)$)
one easily obtains $\beta_1=1$ and then
$c(\beta=1,w^2)=v(\beta_c,w^2)$.
We can distinguish the two cases defined below.
\begin{itemize} 
\item If $v(\beta,w^2)<0$ for some $0<\beta<1$ then we are in the no-error
phase and the turbo decoding algorithm converges to the 
message with ``velocity'' $c(\beta=1,w^2) = \min_{0<\beta<1}v(\beta,w^2)$.
We expect the condition $v(\beta_c,w^2)<0$ to be verified in the 
``low noise'' region $w^2<w^2_{loc}$.
\item If $v(\beta,w^2)\ge 0$ in the interval $0<\beta<1$ then  
$c(\beta=1,w^2)\ge 0$ and the linearization in 
Eq. (\ref{Linearizzata}) is no longer reliable.
In this case $\pi_t(x)$ is expected to converge for $t\to\infty$
to some distribution supported on finite fields $x$. The decoded message
will be plagued by a finite error probability per bit, no matter how many
times do we iterate the turbo decoding algorithm.
\end{itemize}

Let us now study some examples. We consider a gaussian channel
with:
\begin{eqnarray}
P(\hat{J}|\tau) = \frac{1}{(4\pi w^2)^{1/2}}
\exp\left\{-\frac{(\hat{J}-\tau)^2}{4w^2}\right\}\label{Distr1}\\
P(\hat{h}|\tau) = \frac{1}{(2\pi w^2)^{1/2}}
\exp\left\{-\frac{(\hat{h}-\tau)^2}{2w^2}\right\}\label{Distr2}
\end{eqnarray}
This choice of the variances is justified since it corresponds to a 
code with rate $1/3$ (see Ref. \cite{Turbo1}).
It is useful to define the function 
\begin{eqnarray} 
z(\beta,w^2) = \int\! dh\; P(h)\; e^{-2\beta h} = 
\left(\int\! dJ\; P(J)\; e^{-2\beta J}\right)^2=
\exp\left[\frac{2\beta(\beta-1)}{w^2}\right]
\end{eqnarray}
The three cases below have been 
already considered in Ref. \cite{Turbo1}. We refer to the Appendix 
\ref{AppendiceAlgebra} for the calculation of the constants $w_0$
and $w_1$ to be used in Eq. (\ref{PreVelocity2}). 
\renewcommand{\theenumi}{(\alph{enumi})}
\begin{enumerate}
\item A model with nearest neighbours
interaction is: $\epsilon_i(\sg) \equiv \sigma_i\sigma_{i-1}$ and 
$\eta_i(\sg)=\sigma_i$ (which corresponds to the generating polynomials
$g_1(x) = 1+x$ and $g_2(x) = 1$). Using Eq. (\ref{PreVelocity2}) 
and the fact that $w_0(f_1,f_2)=0$ and $w_1(f_1,f_2)=1$
we get
\begin{eqnarray}
v(\beta,w^2) = \frac{1}{\beta}\log \frac{2z^2(\beta,w^2)}{1-z(\beta,w^2)}
\end{eqnarray}
It is easy to see that $v(\beta,w^2)\ge 0$ for each $0<\beta<1$ if
$w^2\ge w^2_{loc} = 1/\log 4$.
\label{Range1}
\item For $\epsilon_i(\sg) \equiv \sigma_i\sigma_{i-1}\sigma_{i-2}$ and 
$\eta_i(\sg)=\sigma_i\sigma_{i-2}$ (generating polynomials:
$g_1(x) = 1+x+x^2$ and $g_2(x) = 1+x^2$) we obtain 
$w_0(f_1,f_2)=2$ and $w_1(f_1,f_2)=2$ and then
\begin{eqnarray}
v(\beta,w^2) = \frac{1}{\beta}\log \frac{2z^5(\beta,w^2)}{1-z^2(\beta,w^2)}
\end{eqnarray}
Finally $w^2_{loc} = -1/(2\log z_c)$ where $z_c$ is the only real solution of 
the equation $2z^5+z^2=1$.
\label{Range3}
\item If we consider the model given by $\epsilon_i(\sg) \equiv 
\sigma_i\sigma_{i-1}\sigma_{i-2}\sigma_{i-3}\sigma_{i-4}$ and 
$\eta_i(\sg)=\sigma_i\sigma_{i-4}$
(generating polynomials:
$g_1(x) = 1+x+x^2+x^3+x^4$ and $g_2(x) = 1+x^4$) we obtain 
$w_0(f_1,f_2)=2$ and $w_1(f_1,f_2)=2$ as in the previous example.
Both $v(\beta,w^2)$ and $w^2_{loc}$ coincide with the ones obtained
above. 
\end{enumerate}

\begin{figure}
\begin{tabular}{c}
\epsfig{figure=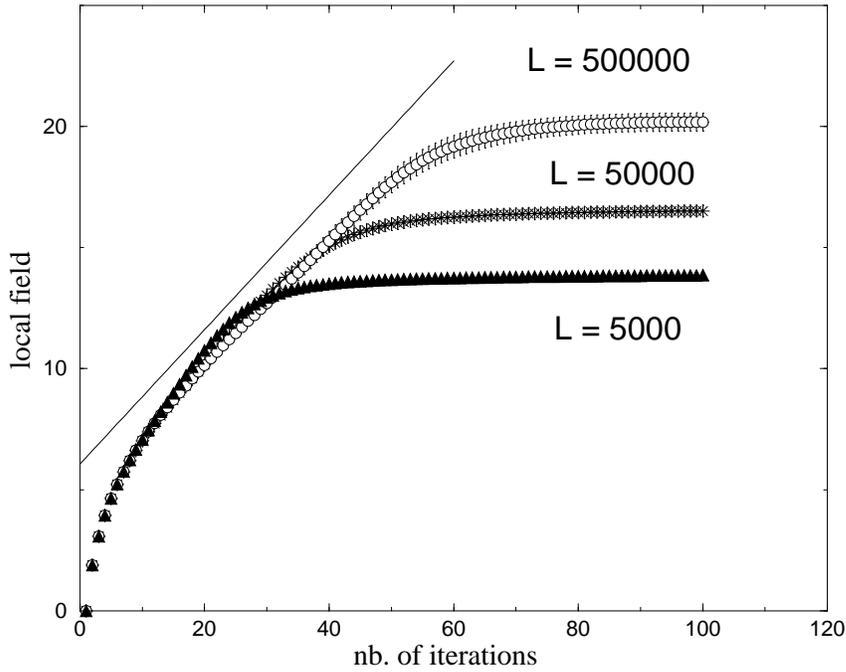,angle=-90,
width=0.7\linewidth}\\
\epsfig{figure=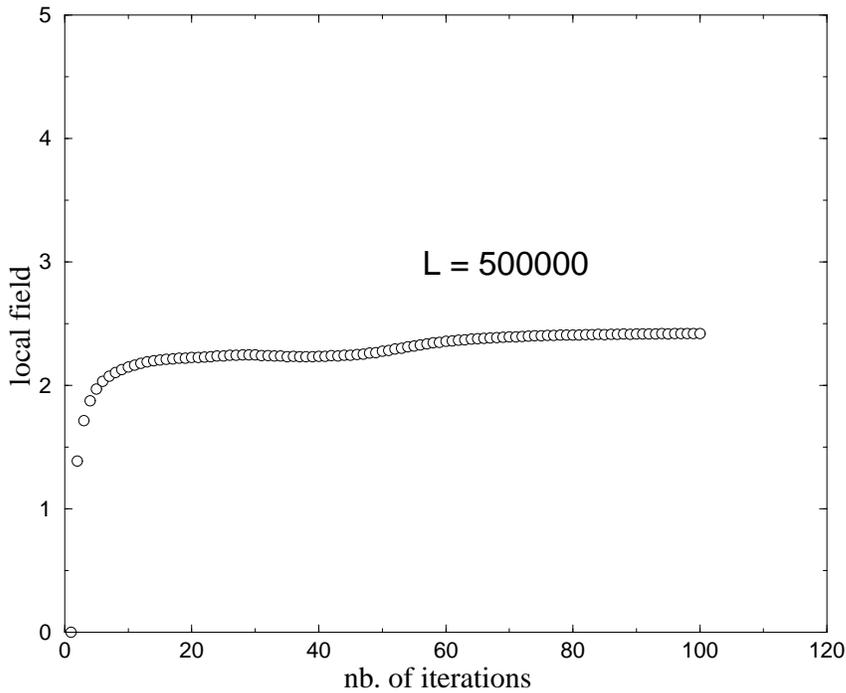,angle=-90,
width=0.7\linewidth}
\end{tabular}
\caption{The dynamics of the turbo decoding algorithm. The graph on
the top gives the average of the local field $\Gamma_i$ (see 
Eqs.(\ref{TurboDecoding1}-\ref{TurboDecoding2})) as a function of the 
number of iterations for different 
sizes of the system. The slope of the straight line 
on the same graph indicates 
the asymptotic velocity obtained in Section \ref{WithoutReplicas2}. 
The graph on the bottom gives the variance of the distribution of the
local field.}
\label{SpeedFig}
\end{figure}

Let us make a few observations about the validity of our calculation. 
The threshold $w^2_{loc}$ has been obtained by starting from a distribution 
$Q(X)$ very near to the ``frozen'' one $Q_*(X)$ and linearizing 
Eq. (\ref{Riscritta}) in $X$. 
It must then be interpreted as a threshold for local 
stability of the ``frozen'' solution.
Moreover, if we take seriously the heuristic derivation of 
Eq. (\ref{Pindependent}),
we can deduce something about the dynamics of the turbo decoding algorithm
in the error-free phase:
the probability distribution of the auxiliary fields $\Gamma_i^{(k)}(t)$ moves
towards infinitely large fields with an average velocity $c(\beta,w^2)$.
This conclusion is compared with numerical data in
Fig. (\ref{SpeedFig}): the agreement seems to be quite good.
An interesting outcome of the previous calculation is that the approach 
to the perfect decoding becomes slower near to the critical signal to
noise ratio.

Let us now discuss the ``non recursive'' models, that is models
such that $g_1(x)$ divides $g_2(x)$. In this case the energy
$\Delta(i)$ to be paid for flipping $\eta_i(\sg)$
remains finite in the thermodynamic limit.
The low $X$ expansions in Eqs. (\ref{LowX1}-\ref{LowX2}) 
have a non vanishing term of order $O(X)$. 
This implies that $Q_*(X)=\delta(X)$ is no longer a
fixed point of Eq. (\ref{Riscritta}).
Let us compute $\Delta(i)$.  
For ``non recursive'' models we can define the polynomial 
$s(x)\equiv\sum_k s_k x^k\equiv g_2(x)/g_1(x)$. 
It is easy to show that
$\Delta(i) = 2J_i+2\sum_k s_k h_{i+k}$. A simple 
approximation of the fixed point of Eq. (\ref{Riscritta}) is:
\begin{eqnarray}
Q_{\infty}(X)\sim\int\!dJ\;P(J)\int\!\prod_{i=1}^w dh_i\;P(h_i)
\delta\left(X-e^{-2\beta J-2\beta\sum_{i=1}^w h_i}\right)
\label{ApprNonRic}
\end{eqnarray}
with $w\equiv{\rm weight}(s)$. This approximation is supposed to be good in 
the low noise region where we expect the distributions $Q_t(X)$ to be 
concentrated on small $X$'s. 
%
%
\section{The replica calculation.}
\label{WithReplicas}
The replica method \cite{SpinGlass} starts with the 
computation of the (integer) moments 
of the partition function. This can be done by introducing an appropriate
order parameter (the choice is a matter of convenience) and by recurring
to standard tricks. Here we choose to use the (multi)-overlaps
$q_{a_1\dots a_l}$ and their complex conjugates
$\hat{q}_{a_1\dots a_l}$: 
\begin{eqnarray}
\overline{Z^n} &=& \int \!\frac{N}{\pi}\;dq_0\;d\hat{q}_0
\int \!\prod_a\frac{N}{\pi}\;dq_a\;d\hat{q}_a
\int \!\prod_{(a,b)}\frac{N}{\pi}\;dq_{ab}\;d\hat{q}_{ab}\;\dots\;
e^{-NS[q,\hat{q}]}\\
S[q,\hat{q}] &=& -1+q_0\hat{q}_0+\sum_a q_a\hat{q}_a
+\sum_{(a,b)} q_{ab}\hat{q}_{ab}+\dots+n\log2+\label{Action}\\
&&+\beta {\cal F}_{1d,n}[q]
+\beta {\cal F}_{1d,n}[\hat{q}]\nonumber\\
{\cal F}_{1d,n}[q]&\equiv& 
-\lim_{N\to\infty}\frac{1}{N\beta} \log Z_{1d,n}[q]\\
Z_{1d,n}[q] &\equiv&\sum_{\{\sigma_i^a\} }\prod_{i=1}^N[q_0+
\sum_a q_a\epsilon_i(\sg^a)+
\sum_{(a,b)}q_{ab}\epsilon_i(\sg^a)\epsilon_i(\sg^b)+\dots]\cdot \\
&&\phantom{\sum_{\sigma_i^a}}
\cdot \int d{\cal P}[\Jg] \exp\left\{-\beta\sum_a H(\sg^a;\Jg)\right\}\nonumber
\end{eqnarray}
where $H(\sg;\Jg) = -\sum_i J_i \epsilon_{i}(\sg)-\sum_i h_i \eta_{i}(\sg)$.
and the replica indices $a,b,\dots$ run from $1$ to $n$.
The usual mean field models have no geometrical structure at all. In
those cases the introduction of the order parameters leads to a 
(replicated) partition function which factorizes over the sites. 
In our case we are left with the problem of computing the one dimensional
partition functions $Z_{1d,n}[q]$. These correspond to the one dimensional 
sub-structures which are not destroyed by the randomness of the model.
The saddle point equations are easily written
\begin{eqnarray}
\hat{q}_{a_1\dots a_l}=\lim_{N\to\infty}\frac{1}{N}\sum_{i=1}^N
\left\<
\frac{\epsilon_i(\sg^{a_1})\cdot \dots\cdot \epsilon_i(\sg^{a_l})}
{[q_0+
\sum_a q_a\epsilon_i(\sg^a)+\dots]}
\right\>_q
\label{Saddle}
\end{eqnarray}
where the expectation values $\<(\cdot )\>_q$, 
$\<(\cdot )\>_{\hat{q}}$ are defined as follows
\begin{eqnarray}
\<(\;\cdot \;)\>_q&\equiv&\frac{1}{Z_{1d,n}[q]}\int d{\cal P}[\Jg]
\sum_{\{\sigma_i^a\} }(\;\cdot \;)\;
\prod_{i=1}^N[q_0+\sum_a q_a\epsilon_i(\sg^a)+\dots]\;
e^{-\beta\sum_a H[\mbox{\boldmath ${\scriptstyle\sigma}$}{}^a;
\mbox{\boldmath ${\scriptstyle J}$}]}
\end{eqnarray}
In the recursive case Eq. (\ref{Saddle}) admits the following solution
\footnote{In fact there is a one parameter family of solutions which are
degenerate. This fact is due to a (not very interesting) symmetry
of the action (\ref{Action}): 
$S[q,\hat{q}]=S[e^{i\theta}q,e^{-i\theta}\hat{q}]$. However 
the integration over the parameter $\theta$ does not pose any problem. We 
shall fix this freedom by imposing $q_0$ to be real.\label{NotaSimmetria}}  
corresponding to a no-error phase: 
$q^*_{a_1\dots a_l}=\hat{q}^*_{b_1\dots b_l}=2^{-n/2}$.
The free energy of this phase is $f_0(\beta) = -2\int dJ\;P(J)\;J
-2\int dh\;P(h)\;h$.
If we parametrize the replica symmetric ansatz as in Ref. \cite{Wong}
\begin{eqnarray}
q_{a_1\dots a_l}=\int_{-\infty}^{+\infty}\!\!dx\;\pi(x)\;\cosh^n(\beta x)\;
\tanh^l(\beta x)&
\end{eqnarray}
and analogously for $\hat{q}_{b_1\dots b_m}$ (with a different distribution 
$\hat{\pi}(x)$), the following free energy functional can is obtained in the
limit $n\to 0$:
\begin{eqnarray}
f[\pi,\hat{\pi}] & = & 
\frac{1}{\beta}\int \!dx\;dy\;\pi(x)\hat{\pi}(y)
\log\left[2\cosh(\beta x+\beta y)\right]+
{\cal F}_{1d}^{RS}[\pi]+{\cal F}_{1d}^{RS}[\hat{\pi}]\\
{\cal F}_{1d}^{RS}[\pi] & \equiv & -\lim_{N\to\infty}
\frac{1}{\beta N}\int \!d{\cal P}[\Jg] \int \!d\pi[\xg]
\log Z_{1d}^{RS}[\Jg,\xg]\\
Z_{1d}^{RS}[\Jg,\xg]&\equiv &\sum_{\sg}\exp\left[
\beta\sum_{i=1}^N ( J_i+x_i) \epsilon_{i}(\sg)
+\beta\sum_{i=1}^N h_i \eta_{i}(\sg)\right]
\end{eqnarray}
The distributions $\pi$ and $\hat{\pi}$ are normalized ( $\int\!dx\;\pi(x)=
\int\! dy\;\hat{\pi}(y)=1$) and satisfy the saddle point equation below:
\begin{eqnarray}
\pi(y) = \lim_{N\to\infty}
\frac{1}{N}\sum_{i=1}^N\int_{-\infty}^{+\infty} 
d\hat{\pi}[\xg ]\int d{\cal P}[\Jg]\ \delta\left(y-\frac{1}{\beta}
\arctanh\left(\<\epsilon_{i}(\sg)\>_{\Jg,\xg}\right)+x_{i}\right)
\label{SaddleRS}
\end{eqnarray}
which is identical to the fixed point equation corresponding to 
Eq. (\ref{Pindependent}), if we suppose the order parameters to be real at
the saddle point.

\begin{figure}
\centerline{
\epsfig{figure=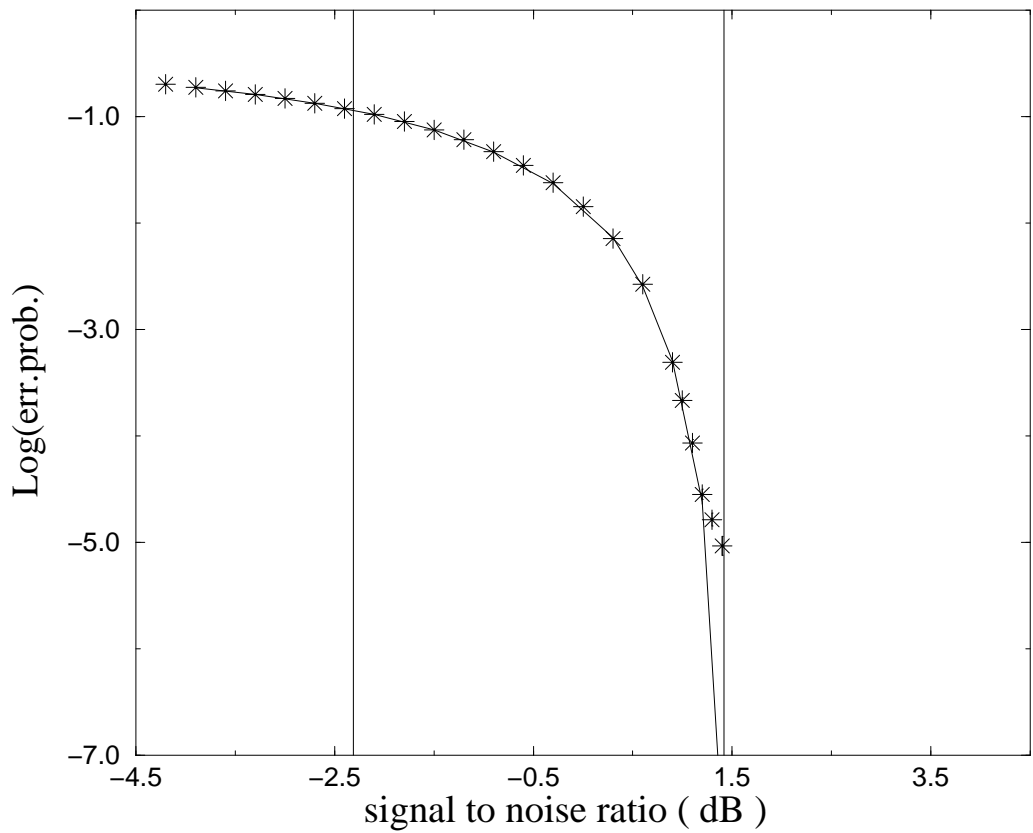,angle=-90,
width=1.0\linewidth}
}
\caption{The numerical results for the error probability per bit 
(stars, $*$), compared with the  analytical prediction (continuous line). 
The analytical prediction is obtained, within the replica symmetric 
approximation, from Eq. (\ref{OmegaRicorsivo}). This graph refers to model
\ref{Range1} defined in Section \ref{WithoutReplicas2}.}
\label{RsaFig}
\end{figure}

Equation (\ref{SaddleRS}) is unpractical since it involves the
unknown distributions $\pi(x)$ and $\hat{\pi}(x)$ infinitely many times. 
However due to the short range structure of the hamiltonians 
defined in Eq. (\ref{Models2}), it can be rewritten as a simple integral 
equation. Obviously the precise form of this equation depends upon the
form of the hamiltonian (\ref{Models2}). In particular it becomes simpler
as the range of the interaction becomes shorter. Let us illustrate this point
by considering the model \ref{Range1} of the previous Section:
$\epsilon_i(\sg)=\sigma_i\sigma_{i-1}$, $\eta_i(\sg)=\sigma_i$.
We start by defining the following (right and left) partition
functions:
\begin{eqnarray}
Z^{(R)}_{i,M}(\sigma_i)& \equiv &\sum_{\sigma_{i+1}\dots\sigma_{i+M}}
\exp\left[\beta\sum_{k=i+1}^{i+M}(J_k+x_k)\sigma_k\sigma_{k-1}+
\beta\sum_{k=i}^{i+M}h_k\sigma_k\right]\\
Z^{(L)}_{i,M}(\sigma_i) & \equiv & \sum_{\sigma_{i-M}\dots\sigma_{i-1}}
\exp\left[\beta\sum_{k=i-M+1}^i(J_k+x_k)\sigma_k\sigma_{k-1}+
\beta\sum_{k=i-M}^i h_k\sigma_k\right]
\end{eqnarray}
and the (right and left) fields:
\begin{eqnarray}
x_i^{R/L} \equiv\lim_{M\to\infty} \frac{1}{2\beta}\log
\frac{Z^{(R/L)}_{i,M}(+)}{Z^{(R/L)}_{i,M}(-)}
\end{eqnarray}
We define now a new couple of order parameters, the 
probability distributions $\omega(x)$
and $\hat{\omega}(x)$ of the right (or left) fields:
\begin{eqnarray}
\omega(x) = \int\!\prod_{i\ge 0}dh_i\;P(h_i)\int\!\prod_{i\ge 1}dJ_i\;P(J_i)
\int\!\prod_{i\ge 1}dx_i\;\pi(x_i)\;
\delta\left(x-x_0^R[J_i;h_i;x_i]\right)
\end{eqnarray}
It easy to show that, at the saddle point, $\omega(x)$ and $\hat{\omega}(x)$
satisfy the following integral equation:
\begin{eqnarray}
\omega(z) &=& \int\!dh\;P(h)\int\!dJ_1\;P(J_1)\int\!dJ_2\;P(J_2)
\int\!dx_1\;\hat{\omega}(x_1)\int\!dx_2\;\hat{\omega}(x_2)
\int\!dz'\;\omega(z')\nonumber\\
&&\phantom{\int\!dh}\delta\left\{z-h
-\Theta_{\beta}\left[z';J_1+J_2
+\Theta_{\beta}(x_1;x_2)\right]\right\}\label{OmegaRicorsivo}\\
\Theta_{\beta}(x;y)&\equiv&\frac{1}{\beta}
\arctanh[\tanh(\beta x)\tanh(\beta y)]
\end{eqnarray} 
and that the solution of Eq. (\ref{SaddleRS}) is related to the solution of 
the previous equation as follows:
\begin{eqnarray}
\pi(x) = \int\!dJ\;P(J)\int\!dx_L\;\hat{\omega}(x_L)
\int\!dx_R\;\hat{\omega}(x_R)\;
\delta[x-J-\Theta_{\beta}(x_R;x_L)]
\end{eqnarray}
Equation (\ref{OmegaRicorsivo}) reduces to the Dyson Schmidt equation 
\cite{Dyson,Schmidt,Luck} for a one dimensional model with nearest neighbour 
interaction if we keep the distribution $\hat{\omega}(x)$ fixed. 
The interaction between the two one-dimensional subsystems turns 
it into a nonlinear equation. Moreover Eq. (\ref{OmegaRicorsivo}) can be 
treated numerically more easily than Eq. (\ref{SaddleRS}). A possible approach
consists in representing the unknown distribution as $\omega(x) = 
\sum_{j=1}^{K}\delta(x-x_j)$ and iterating Eq. (\ref{OmegaRicorsivo}) until
a fixed point is reached. An example of this kind of computations is
shown in Fig. (\ref{RsaFig}). 

It is simple to obtain the analogous
of Eq. (\ref{OmegaRicorsivo}) for the simplest non recursive model, defined
by: $\epsilon_i(\sg)=\sigma_i$, $\eta_i(\sg)=\sigma_i\sigma_{i-1}$. 
The final result is
\begin{eqnarray}
\omega(z) &=& \int\!dh\;P(h)\int\!dJ_1\;P(J_1)\int\!dJ_2\;P(J_2)
\int\!dx_1\;\hat{\omega}(x_1)\int\!dx_2\;\hat{\omega}(x_2)
\int\!dz'\;\omega(z')\nonumber\\
&&\phantom{\int\!dh}\delta\left(z-
\Theta_{\beta}\left[h;J_1+J_2+
x_1+x_2+z'\right]\right)\label{OmegaNonRicorsivo}\\
\pi(x) &=& \int\!dJ\;P(J)\int\!dx_L\;\hat{\omega}(x_L)\int\!dx_R\;
\hat{\omega}(x_R)\;\delta(x-J-x_L-x_R)
\end{eqnarray} 
A simple approximation to the solution of Eq. (\ref{OmegaNonRicorsivo}) can
be obtained by starting from a distribution $\omega(x)$ supported on very
large fields $x$ and iterating Eq. (\ref{OmegaNonRicorsivo}) one time.
The result is $\pi(x)\sim \int\!dJ\;P(J)\int\!dh_1\;P(h_1)\int\!dh_1\;
P(h_1)\;\delta(x-J-h_1-h_2)$, which coincides with the 
more general Eq. (\ref{ApprNonRic}) after the change of variables 
$X=e^{-2\beta x}$. No such approximation 
is possible for Eq. (\ref{OmegaRicorsivo}). 

Expressions equivalent to Eqs. (\ref{OmegaRicorsivo}-\ref{OmegaNonRicorsivo})
can be derived for more complicated
types of interaction. In general the distribution $\omega(x)$, which is
defined on the real line, will be replaced by a distribution defined on 
$\mathbb{R}^{2^r-1}$, $r$ being the range of the hamiltonian.
%
%

\section{The stability of the frozen solution.}
\label{Stability}
We would like to study local stability of the no-error phase in the context
of the replica method. This can be done\footnote{For similar calculations
see Ref. \cite{Motti}.} by computing the eigenvalues
of the matrices:
\begin{eqnarray}
M^\pm _{a_1\dots a_l,b_1\dots b_m}[q] = \delta_{a_1\dots a_l,b_1\dots b_m}
\pm \frac{\partial^2 \beta {\cal F}_{1d,n}[q]}
{\partial q_{a_1\dots a_l}\partial q_{b_1\dots b_m}}
\end{eqnarray}
$M^\pm [q]$ are the mass matrices for purely real ($M^+[q]$), or
purely immaginary ($M^-[q]$), fluctuations of the order parameter around the
value $q$. We are interested in the saddle point 
$q^*_{a_1\dots a_l}= 2^{-n/2}$.
In order to write down all the $2^n$ eigenvectors of 
$M^\pm [q^*]$ it is convenient to change slightly our notation for the 
overlaps. Let us denote by $\Omega\subset \{1,2,\dots,n\}$ the set of 
$l\equiv |\Omega|$ different indices $(a^{\Omega}_1,\dots,a^{\Omega}_l)$.
We can use the $\Omega$'s as indices for the overlaps with the natural 
identification $q_{\Omega}\equiv q_{a^{\Omega}_1\dots a^{\Omega}_l}$.
It is not difficult to show that
\begin{eqnarray}
T^{(N)}_{\Omega_a,\Omega_b} 
&\equiv& \left.\frac{1}{N}\frac{\partial^2\log Z_{1d,n}[q]}
{\partial q_{a_1\dots a_l}\partial q_{b_1\dots b_m}}\right|_{q=q^*}=\\
&=&\frac{2^{1-n}}{N}\left\{\sum_{(i,j)}
\int\!d{\cal P}[\Jg]\;e^{nN\beta f-n\beta E_0}\;
\left(1+e^{-\beta\Delta(i,j)}\right)^n
\left[
\tanh\left(\frac{\beta\Delta(i,j)}{2}\right)
\right]^{u}-\frac{N^2}{2}\right\}\nonumber
\end{eqnarray}
where $\Delta(i,j)$ is defined in Section \ref{WithoutReplicas2},
$e^{-nN\beta f}\equiv\int\!d{\cal P}[\Jg]\;e^{-n\beta H(\sg_0)}$
and $u\equiv u_{a_1\dots a_l,b_1\dots b_m}$ counts the indices
which are either in the set in $\Omega_a\equiv(a_1,\dots,a_l)$ or in the set
$\Omega_b\equiv(b_1,\dots,b_m)$ but not in both.
If $q$ is an eigenvector of $T^{(N)}$ with eigenvalue $\theta_N$,
then it is an eigenvector of  $M^\pm [q^*]$ with eigenvalue
$\mu ^\pm  = 1\mp\lim_{N\to\infty}\theta_N$.

Notice that $T^{(N)}$ is an hermitian matrix with
respect to to the scalar product:
\begin{eqnarray}
\<q,q'\>_n\equiv \sum_{l=0}^n \sum_{(a_1,\dots,a_l)} 
q^*_{a_1\dots a_l}q'_{a_1\dots a_l}=\sum_{\Omega} q^*_{\Omega}q'_{\Omega}
\label{ScalarProduct}
\end{eqnarray}
We shall use another subset of $\{1,\dots,n\}$ (let us call it $\Lambda$) 
to label the different eigenvectors of $T^{(N)}$, which we now exhibit:
\begin{eqnarray}
q_{\Omega}^{(\Lambda)} \equiv \frac{1}{2^{n/2}}(-1)^{|\Lambda\cap\Omega|}
\end{eqnarray}
The vectors $\{ q^{(\Lambda)}\}$ form an orthonormal set with respect
to the scalar product defined in Eq. (\ref{ScalarProduct}).
This is easily proven by induction on $n$.
The vector $q^{(\varnothing)}$ is nothing but the constant one. 
The corresponding eigenvalue is 
$\theta^{(\varnothing)}_N=-1$, whence $\mu _{(\varnothing)}^+ = 2$ and 
$\mu _{(\varnothing)}^- = 0$. The eigenvalue $\mu _{(\varnothing)}^- = 0$ is 
a remnant of the invariance of the action under the symmetry cited in the 
footnote \ref{NotaSimmetria} of the previous Section.
In order to compute the eigenvalues in the subspace orthogonal to 
$q^{(\varnothing)}$, the following formula turns out to be useful:
\begin{eqnarray}
\sum_{\Omega'} x^{|\Omega\bigtriangleup\Omega'|}\;q^{(\Lambda)}_{\Omega'} 
= (1-x)^{|\Lambda|}(1+x)^{n-|\Lambda|}\; q^{(\Lambda)}_{\Omega}
\label{Evalue1}
\end{eqnarray}
where $\Omega\bigtriangleup\Omega'$ denotes the symmetric difference
of $\Omega$ and $\Omega'$ (i.e. $\Omega\bigtriangleup\Omega'\equiv
(\Omega\setminus \Omega')\cup(\Omega'\setminus \Omega)$).
Using Eq. (\ref{Evalue1}) and the results of algebra outlined in 
Section \ref{WithoutReplicas2} we get (for $\Lambda\neq\varnothing$):
\begin{eqnarray}
\theta_{N\to\infty}(\Lambda) = 2 \zeta_J^2
\sum_{m=1}^{\infty}\zeta_h^{\mathrm{weight}(s_m)}
\label{Evalue2}
\end{eqnarray}
where ${\rm weight}(s_m)$ is defined in Section \ref{WithoutReplicas2} and
\begin{eqnarray}
\zeta_C = \zeta_C (|\Lambda|,n,\beta) = 
\frac{\displaystyle\int\!dC\;P(C)\;e^{(n-2|\Lambda|)\beta C} }
{\displaystyle\int\!dC\;P(C)\;e^{n\beta C}}
\end{eqnarray}
for $C\to h$ or $C\to J$. When the one dimensional hamiltonians 
(\ref{Models2}) satisfy the hypothesis of Lemma \ref{Eccitazioni2energia},
the sum in Eq. (\ref{Evalue2}) can be explicitly computed yielding:
\begin{eqnarray}
\theta_{N\to\infty}(\Lambda) = 
\frac{2\;\zeta_J^2 \;\zeta_h^{w_0(f_1,f_2)+w_1(f_1,f_2)} }
{1-\zeta_h^{w_1(f_1,f_2)}}
\label{Evalue3}
\end{eqnarray}
If $n\ge 2|\Lambda|$ then $\theta(\Lambda,n;\beta)$ is positive 
and decreasing with 
$\beta$. Moreover $\lim_{\beta\to\infty} \theta(\Lambda,n;\beta)=0$ and
$\lim_{\beta\to 0} \theta(\Lambda,n;\beta)=\infty$. 
We can thus define the critical temperatures $\beta_{l,n}$
for $n/2\ge l=|\Lambda|\ge 1$, by requiring
\footnote{Notice that Eq. (\ref{Critical}) can have more than one solution 
for $n<2|\Lambda|$. The ``physical'' critical point is obtained by
taking the limit $n\to 0$ of the solution of Eq. (\ref{Critical})
which exists for any $n$.} 
\begin{eqnarray}
\theta(\Lambda,n;\beta_{|\Lambda|,n}) = 1
\label{Critical}
\end{eqnarray}
If $\beta >\beta_{|\Lambda|,n}$ the ``frozen'' saddle point is stable
with respect to the direction $q^{(\Lambda)}$.  
If $\beta <\beta_{|\Lambda|,n}$ it becomes unstable: 
$\mu ^+_{(\Lambda)}=1-\theta(\Lambda)<0$ while 
$\mu ^-_{(\Lambda)}=1+\theta(\Lambda)>0$ (it could be guessed 
that the ``imaginary'' directions would be stable because 
of the physical interpretation of the overlaps).
In the limit $n\to 0$, $\beta_{l,n}\to \beta_c/l$: the critical directions 
are the ones corresponding to $|\Lambda|=1$. It is easy to see that the 
critical temperature $\beta_c$ coincides with the one obtained in 
Section \ref{WithoutReplicas2}.
%
%

\section{Conclusion.}

\label{Conclusion}

We have presented two derivations of the local stability condition for the 
no-error phase. Both will be object of the criticism of the skeptical reader.
In the first one we obtained the ``mean field'' equation describing the 
dynamics of the decoding algorithm, Eq. (\ref{Pindependent}),
by making use of heuristic arguments. Indeed we argued 
Eq. (\ref{Pindependent}) to be valid only in the replica symmetric 
approximation. 
In the second derivation we made use of the replica 
method, which has not (yet) well founded mathematical basis.

We think that the two derivations compensate each other for their defects. 
Moreover they yield 
the same replica symmetric saddle point equation (\ref{SaddleRS})
and give the same picture of the instability which destroys the
no-error (frozen) phase.
This corresponds to couples of flipped $\epsilon(\sg)$'s. 
Finally thanks to the first derivation we get some insight on the 
behavior of the decoding algorithm. In particular we have seen
that, in the frozen phase, it approaches a no-error fixed point. 
This approach becomes slower near to the boundary of the frozen phase. 

In Ref. \cite{Turbo1} the local stability threshold computed here has been 
compared with numerical simulations for two types of code, respectively
the models \ref{Range1} and \ref{Range3} presented in Section 
\ref{WithoutReplicas2}. Good agreement was found only for model 
\ref{Range1}. We propose two possible explanations of the disagreement 
for model \ref{Range3}:
\begin{itemize}
\item the phase transition is a first order one;
\item the turbo decoding algorithm used in Ref. \cite{Turbo1} gets sticked
in some local minimum of the free energy, characterized by a finite error
probability per bit.
\end{itemize}
We have not yet enough informations for choosing between these two scenarios.
%
%

\appendix

\section{Useful algebra results.}
\label{AppendiceAlgebra}
\newtheorem{teoremiappendice}{Theorem}[section]
\newtheorem{lemmiriciclati}{Lemma}[section]

In this Appendix we remind to the reader some known facts in 
the theory of finite fields and we prove the propositions stated in Section
\ref{WithoutReplicas2}. These are nothing but simple exercises and we 
work out them in detail only for greater convenience of the reader.
Finally we illustrate a few applications of the results obtained.
The reader interested in a more complete treatment can consult
Refs. \cite{Roman1, Roman2}.
 
Let us begin with some elementary definitions. 
The basic object is $\mathbb{Z}_2$ i.e. the field of integer numbers modulo 
$2$. A polynomial over  $\mathbb{Z}_2$,  $f(x)\in\mathbb{Z}_2[x]$ 
is simply a polynomial whose coefficients are in $\mathbb{Z}_2$. 
We say $f(x)\in\mathbb{Z}_2[x]$ to be irreducible if 
there do not exist two noncostant polynomials $g(x),h(x)\in\mathbb{Z}_2[x]$
such that $f(x) = g(x)\cdot h(x)$.
Any $f(x)\in\mathbb{Z}_2[x]$ possess an unique factorization, i.e. a 
decomposition of the form $f(x) = f_1(x)^{r_1}\cdot \dots\cdot f_h(x)^{r_h}$
where $f_i(x)\in\mathbb{Z}_2[x]$ are irreducible and $r_i\ge 1$ are integer
numbers. Given two polynomials $f(x),g(x)\in\mathbb{Z}_2[x]$
we say that $f(x)$ divides $g(x)$ (in symbols $f(x)|g(x)$) if there 
exists $h(x)\in \mathbb{Z}_2[x]$ such that $g(x) = f(x)\cdot h(x)$
\footnote{Similarly, given two integer numbers $p,q\in\mathbb{Z}$,
we say that $p$ divides $q$ (and write $p|q$) if there exists $m\in\mathbb{Z}$,
such that $q=mp$.}.
For an irreducible polynomial $f(x)\in\mathbb{Z}_2[x]$ it does make sense
to define the order $o(f)$: $o(f)$ is the smallest positive integer $k$ such 
that $f(x)|x^k+1$. The basic result which we shall employ in this Appendix is 
the following: 
\begin{teoremiappendice}
\label{Irriducibili}
Let $f(x)$ be an irreducible polynomial over $\mathbb{Z}_2$.
Then $f(x)|x^k+1$ if and only if $o(f)|k$. 
\end{teoremiappendice}
It is useful to know how to compute the order of an irreducible polynomial.
The main tool is the theorem below:
\begin{teoremiappendice}
\label{Ordine}
Let $f(x)$ be an irreducible polynomial of degree $d$ over $\mathbb{Z}_2$.
Then $d$ is the smallest positive integer for which $o(f)|2^d-1$.
\end{teoremiappendice}
Moreover it is obvious from the definition that $o(f)\ge\deg(f)$

Our first step will be the proof of
Lemma \ref{Eccitazioni2possibili} which we restate here as follows
\begin{lemmiriciclati}
\label{Eccitazioni2possibilibis}
Let $f(x)$ be a polynomial on $\mathbb{Z}_2$ with the following factorization
\begin{eqnarray}
f(x) = f_1^{r_1}(x)\cdot \dots\cdot f_h^{r_h}(x)\quad ;\quad\quad
r_i\ge 1
\end{eqnarray}
where the polynomials $f_i(x)$ are irreducible over $\mathbb{Z}_2$.
Let $p_i$ be the smallest integer such that $2^{p_i}\ge r_i$.
Then $f(x)|(1+x^k)$ if and only if $2^{p_i}|k$ and $o(f_i)|k$ for 
$i\in\{1,\dots,h\}$.
\end{lemmiriciclati}
{\it Proof of Lemma \ref{Eccitazioni2possibilibis}.} Let us begin by noticing
that, since the $f_i(x)$ are irreducible, $f(x)|(1+x^k)$ if and only
if $f_i^{r_i}(x)|(1+x^k)$ for $i\in\{1,\dots,h\}$. We can then 
limit ourselves to the case $f(x)=h^r(x)$ with $h(x)$ irreducible.
It is convenient to work in an extension of $\mathbb{Z}_2$, i.e. in a field 
containing $\mathbb{Z}_2$ as a subfield. We choose an extension 
(let us call it $S$)
of $\mathbb{Z}_2$ such that both $h(x)$ and $(1+x^k)$ can be decomposed in
linear factors. The existence of such an extension is a basic fact of field
theory. We are then looking for the $k$ such that all the root 
of $h(x)$ (in $S$) are roots of $(1+x^k)$ with multiplicity at least $r$.
It is then necessary to study the multiplicity of the roots of
$(1+x^k)$. The first observation is that, if $k$ is odd, all the roots are
simple. In fact $\frac{d}{dx}(1+x^k)=kx^{k-1}$ has no roots in common with
$(1+x^k)$. The second observation consists in noticing that 
$(1+x^{2k})=(1+x^k)^2$. We deduce that $(1+x^{2^mk})$ with $k$ odd
has $k$ distinct roots (the same as $(1+x^k)$), each one with multiplicity
$2^m$. The final outcome is that $h^r(x)|(1+x^{2^mk})$ if and only if
$2^m\ge r$ and $o(h)|k$ $\square$
             
From Lemma \ref{Eccitazioni2possibilibis} the explicit form of the period
$\omega(f)$ used in Sec. \ref{WithoutReplicas2} is easily obtained:
\begin{eqnarray}  
\omega(f) = 2^{\max (p_1,\dots,p_h)}\;{\rm lcm}(o(f_1),\dots,o(f_h))
\end{eqnarray}
The Lemmas \ref{Eccitazioni1} and \ref{Eccitazioni2energia} are easy 
consequences of Lemma \ref{Eccitazioni2possibili}. 

{\it Proof of Lemma \ref{Eccitazioni1}.} Let us begin by 
considering the series $1/f(x)$. We can always define
the polynomials $\varphi_n(x)$ with $\deg(\varphi_n)<\omega(f)$ 
such that $1/f(x) = \sum_{n=0}^{\infty}\varphi_n(x)\; x^{n\omega(f)}$.
Since $f(x)$ divides $(1+x^{m\omega(f)})$ for all $m\ge 1$, we conclude that
$\varphi_n(x)=\varphi_{n'}(x)\equiv\varphi(x)$ for all $n,n'\ge 0$ and
$(1+x^{m\omega(f)})/f(x) = \sum_{k=0}^{m-1} \varphi(x)\;x^{k\omega(f)}$.
With the following definition
\begin{eqnarray}
g(x)\varphi(x)\equiv\sum_{l=0}^L g_l(x)\;x^{l\omega(f)}\quad,\quad
{\rm deg}[g_l(x)]<\omega(f)
\label{Prodotto}
\end{eqnarray}
we get
\begin{eqnarray}
\frac{g(x)}{f(x)} = \sum_{n=0}^{\infty}x^{n\omega(f)}
\sum_{l=0}^{\min(n,L)}g_l(x) \equiv \sum_{n=0}^{\infty}x^{n\omega(f)} p_n(x)
\end{eqnarray}
Notice that, for $n\ge L$, $p_n(x) = p_{\infty}(x) \equiv g(x)/f(x) \bmod 
x^{\omega(f)}$. An upper bound on $L$ is easily obtained from
Eq. (\ref{Prodotto}) yielding\footnote{Here use the definition 
$\lceil x\rceil \equiv\min\{n\in\mathbb{Z}\; :\; n>x\}$.}  
$p_n(x) = p_{\infty}(x)$ for
$n\ge \lceil(\deg[g(x)]-1)/\omega(f)\rceil\ge L$. 
Clearly it cannot be  $p_{\infty}(x)=0$ otherwise we would 
conclude that $f(x)$ divides $g(x)$ in contradiction with the hypothesis. In 
order to complete the proof of
let us suppose the following equation to hold
\begin{eqnarray}
\frac{g(x)}{f(x)} = \sum_{n=0}^{\infty}x^{n\omega'} p'_n(x)
\end{eqnarray}
with $\omega'<\omega(f)$, $\deg(p'_n)<\omega'$ and $p'_n(x) = p'_{\infty}(x)$
for $n$ large enough. This implies that $f(x)$ divides $g(x)(1+x^{\omega'})$
but, since $\gcd(f,g)=1$, we would conclude that $f(x)$ divides 
$(1+x^{\omega'})$ contradicting Lemma \ref{Eccitazioni2possibili} $\square$ 

{\it Proof of Lemma \ref{Eccitazioni2energia}.} It suffices to specialize
the content of the previous paragraph to the case $\deg[g(x)]\le\omega(f)$:
\begin{eqnarray}
g(x)\varphi(x) &= &g_0(x)+g_1(x)\;x^{\omega(f)}\label{GG}\\
s_m(x)\equiv\frac{g(x)}{f(x)}(1+x^{m\omega(f)}) &=&
g_0(x)+\{g_0(x)+g_1(x)\}\sum_{h=1}^{m-1}x^{h\omega(f)}+g_1(x)
\end{eqnarray}
whence
\begin{eqnarray}
\mbox{weight}[s_m(x)] &=& w_0+ w_1\cdot m
\label{Peso1}\\
w_0&\equiv& \mbox{weight}[g_0(x)]+\mbox{weight}[g_1(x)]
-\mbox{weight}[g_0(x)+g_1(x)]
\label{Peso2}\\
w_1&\equiv& \mbox{weight}[g_0(x)+g_1(x)]
\label{Peso3}
\end{eqnarray}
$\square$

What does it happen when the hypothesis of Lemma \ref{Eccitazioni2energia}
are not satisfied?
It is easy to guess the answer. There exists a positive integer $m_0$
such that, for $m\ge m_0$, ${\rm weight}(s_m)$ grows linearly with $m$:
${\rm weight}(s_m) = \tilde{w}_0(f,g)+\tilde{w}_1(f,g)\cdot m$
with $\tilde{w}_1 (f,g)= {\rm weight}(p_{\infty})$. 
Thanks to this fact we can always sum the series in Eq. (\ref{PreVelocity1})
in the interval $0<\beta<\beta_1$. The discussion of the behavior of 
Eq. (\ref{Linearizzata}) presented in Section \ref{WithoutReplicas2} is 
then completely general. 

Let us return down to the earth and make a few examples. We shall
consider the codes presented in Ref. \cite{Turbo1}:
\renewcommand{\theenumi}{(\alph{enumi})}
\begin{enumerate}
\item The simplest non trivial case: $f(x) = 1+x$, $g(x) = 1$. Clearly
both the polynomials are irreducible. The degree of $f(x)$ is 
$\deg[f(x)]=1$. Because of Theorem \ref{Ordine} $o(f)|2^1-1=1$ whence 
$o(f)=1=\omega(f)$. Theorem \ref{Irriducibili} implies that 
$f(x)|1+x^k$ for each $k\ge 1$. This conclusion is easily 
confirmed by the well known formula 
$(1+x^k)=(1+x)(1+x+\dots +x^{k-1})$. Lemma \ref{Eccitazioni1} tells us that
$g(x)/f(x) = \sum_{n=0}^{\infty}p_nx^n$ with $p_n=p_{\infty}$ for $n\ge 0$
and that $p_{\infty}=1$ ($1$ is the unique non zero polynomial of degree zero).
We have thus rediscovered the simple fact that 
$(1+x)^{-1}=\sum_{n=0}^{\infty}x^n$. Finally we observe the hypothesis of
Lemma \ref{Eccitazioni2energia} are satisfied and that (with the notation
of Eq. (\ref{GG})), $g_0(x) = 1$ and $g_1(x) = 0$. 
From Eqs. (\ref{Peso1}-\ref{Peso2}-\ref{Peso3}) it follows that 
$\mbox{weight}[s_m(x)=(1+x^m)/(1+x)] = m$ 
which is easily confirmed by observing that 
$s_m(x) = 1+x+\dots+x^{m-1}$.
\label{CodiceSemplice}
\item A less elementary example is: $f(x)=1+x+x^2$, $g(x) = 1+x^2$. 
It is easy to see that $f(x)$ is irreducible and that $g(x) = (1+x)^2$
whence $\gcd(f,g)=1$.
From $o(f)|2^{\deg(f)}-1=3$ and $o(f)\ge\deg(f)=2$ we deduce that 
$o(f)=3=\omega(f)$. In fact
\begin{eqnarray}
\frac{1}{1+x+x^2} & = & 1+x+x^3+x^4+x^6+\dots = 
\sum_{n=0}^{\infty} \varphi(x)x^{3n}\\
\varphi(x) & = & 1+x
\end{eqnarray}
Thus by Lemma \ref{Eccitazioni2possibili} 
$f(x)|(1+x^k)$ if and only if $k$ is a multiple of $3$. 
We can use Lemma \ref{Eccitazioni2energia} in order to compute
the weight of $s_m(x)=(1+x^2)(1+x^{3m})/(1+x+x^2)$.
We see that $g_0(x)=1+x+x^2$ and $g_1(x) = 1$ whence 
$\mbox{weight}[h_m(x)]=2+2m$. With some book-keeping one can confirm
this result:
\begin{eqnarray}
h_m(x)=
x+\sum_{l=0}^{m-1}(x^{3l+1}+x^{3l+2})+x^{3m}
&\to& \mbox{weight}[h_m(x)]=2+2m
\end{eqnarray}
\item Finally the generating polynomials used in Ref. \cite{PrimoBerrou} to 
build the first example of turbo code: $f(x) = 1+x+x^2+x^3+x^4$, 
$g(x) = 1+x^4$. Also in this case $f(x)$ is irreducible and 
$g(x) = (1+x)^4$ yielding $\gcd(f,g)=1$. Since $o(f)|2^{\deg(f)}-1=15$
and $o(f)\ge \deg(f)=4$, we deduce that either $o(f)=5$ or $o(f)=15$.
However we know that $(1+x^5)=(1+x)(1+x+x^2+x^3+x^4)$ and we conclude that
$o(f)=5=\omega(f)$. In fact
\begin{eqnarray}
\frac{1}{1+x+x^2} & = & 1+x+x^5+x^6+x^{10}+x^{11}+\dots = 
\sum_{n=0}^{\infty} \varphi(x)x^{5n}\\
\varphi(x) & = & 1+x
\end{eqnarray} 
Using the fact that $g_0(x) = 1+x+x^4$ and $g_1(x) = 1$ we get 
$\mbox{weight}[h_m(x)]=2m+2$.
\end{enumerate}
%
%

\section{On the asymptotic behavior of the solutions of  
Eq. (\ref{Linearizzata}).}
\label{AppendiceProbabilita}

In this Appendix we study Eq. (\ref{Linearizzata}) in order to extend to
this case the results concerning Eq. (\ref{Alberi}) used in Section
\ref{WithoutReplicas2}. We shall examine both the approach of Ref. 
\cite{DerridaSpohn}, which is based on the analogy with the KPP equation
and is a non rigorous one, and the approach of Ref. \cite{BuffetTrees},
which employs probability theory and is entirely satisfactory from the 
mathematical point of view.

We would like to deal with this type of equation:
\begin{eqnarray}
Q_{n+1}(Z) & = &\int_{-\infty}^{\infty}\!P(V)\;dV
\int_{-\infty}^{\infty}\prod_{i=1}^{\infty}p(h)\;dh
\int_0^{\infty}\prod_{i=1}^{\infty}Q_n(Z_i)\;dZ_i\nonumber\\
&&\quad\delta\left(Z-\sum_{i=1}^{\infty}\exp\left\{-\beta V
-\beta\sum_{j=1}^{i-1}h_j\right\}Z_i\right) 
\label{Type}
\end{eqnarray}
with the requirement that $\int\!dh\;p(h)\;h>0$ and the initial
condition $P_0(Z)=\delta(Z-1)$. Following Ref.
\cite{DerridaSpohn} we introduce the function:
\begin{eqnarray}
G_n(x) \equiv \int_0^{\infty} \!dZ\;Q_n(Z) \exp\{- e^{-\beta x} Z\} 
\label{DefinizioneGen}
\end{eqnarray} 
which satisfy this recurrence equation
\begin{eqnarray}
G_{n+1}(x) = \int_{-\infty}^{\infty}\!P(V)\;dV
\int_{-\infty}^{\infty}\prod_{i=1}^{\infty}p(h_i)\;dh_i\;
\prod_{i=1}^{\infty}G_n(x+V+\sum_{j=1}^{i-1}h_j)
\label{RecurrenceGen}
\end{eqnarray}
Let us make a few elementary observations concerning Eq. (\ref{RecurrenceGen}):
if $0\le G_m(x)\le 1$ for some $m$ and all $x$ then $0\le G_n(x)\le 1$ for all
$x$ and $n>m$; if $\lim\sup_{x\to\infty}G_n(x) = g_{\infty}<1$ 
then $G_{n+1}=0$; if $G_n(x)$ is increasing and $0<G_n(x)<1$ for some $x$
(both these hypothesis are implied by Eq. (\ref{DefinizioneGen})) then
$\lim_{x\to -\infty}G_{n+1}(x) =0$.
The stationary uniform solutions of Eq. (\ref{RecurrenceGen}) are 
$G^A_n(x) = 0$ and $G^B_n(x) = 1$. The first one is obviously stable. 
If we consider a small fluctuation around $G^B_n(x)$, 
$G_n(x) = 1+\rho_n(x)$ we get:
\begin{eqnarray}
\rho_{n+1}(x) \simeq \int_{-\infty}^{\infty}\!P(V)\;dV
\int_{-\infty}^{\infty}\prod_{i=1}^{\infty}p(h)\;dh\;
\sum_{i=1}^{\infty}\rho_n(x+V+\sum_{j=1}^{i-1}h_j)
\end{eqnarray}
which can be diagonalized in Fourier space: 
\begin{eqnarray}
{\hat \rho}_{n+1} (k) \simeq \frac{\hat{P}(k)}{1-\hat{p}(k)}
{\hat \rho}_n (k)\equiv \lambda(k)
{\hat \rho}_n (k)
\end{eqnarray}
Notice that $|\lambda(k)|>1$ for $k$ small enough and that $|\lambda(k)|$
diverges at $k=0$ in agreement with the previous observation that 
if $G_n(x) = 1-\rho$ than $G_{n+1}(x)=0$. 
The preceding observations lead us to the hypothesis that the
$n\to\infty$ behavior of our problem is controlled by front-like solutions
$G_n(x) = g(x-c(\beta)n)$ interpolating between the stable state 
$G^A_n(x) = 0$ at $x\to -\infty$ and $G^B_n(x) = 1$ at $x\to \infty$.

This scenario is easily confirmed in the case without disorder. If
$P(V) = \delta(V-V_0)$ and $p(h) = \delta(h-h_0)$ one obtains
$P_n(Z) = \delta(Z-e^{\beta c(\beta)n})$, 
$G_n(x) = \exp\{-e^{-\beta(x-c(\beta)n)}\}$ with
\begin{eqnarray}
c(\beta) = \frac{1}{\beta}\log\frac{e^{-\beta V_0}}{1-e^{-\beta h_0}}
\end{eqnarray}
In the general case we assume the existence of front-like solutions with 
the large $x$ behavior $G_n(x)\sim
1-e^{-\beta(x-c(\beta)n)}+o(e^{-\beta x})$.
The front velocity is obtained through the construction given in 
Eqs. (\ref{FrontVelocity}-\ref{CriticalTemperature}) with
\begin{eqnarray}
v(\beta) \equiv \frac{1}{\beta}\log\phi(\beta) =
\frac{1}{\beta}\log\frac{\<e^{-\beta V}\>}{1-\<e^{-\beta h}\> }
\label{TreeSpeed}
\end{eqnarray}
Notice that $\<h\> >0$ implies that $\<e^{-\beta h}\> <1$ 
in some interval $0<\beta<\beta_1$ and that $\beta_c<\beta_1$. 
This remark allows us to sum the series $\sum_k \<e^{-\beta h}\>^k$ in the 
range $0<\beta<\beta_c$, thus obtaining Eq. (\ref{TreeSpeed}). 
The same remark will be useful in the following.

Let us consider now the more rigorous approach used in Ref. \cite{BuffetTrees}.
We start by defining the polymer model which corresponds to 
Eq. (\ref{Type}). We have to use a tree with a numerable set of branches 
stemming from each node. A node of the $n$-th generation is identified 
by $n$ integer numbers $\underline{\omega} \equiv (\omega_1,\dots,\omega_n)$;
its generation is denoted by $|\underline{\omega}|$.
We denote by $\underline{0}$ the root node (i.e. the only node of the zeroth
generation).
We say that the node $\underline{\omega}'$ belonging to the $m$-th generation
is a descendant of the node 
$\underline{\omega}$ of the $n$-th generation
(and write $\underline{\omega}\prec\underline{\omega}'$ if $n<m$ or
$\underline{\omega}\preceq\underline{\omega}'$ if $n\le m$ )
if $\omega_1 = \omega'_1,\dots,\omega_n=\omega'_n$. 
The node $\underline{\omega}'$
is said to be an older brother of the node 
$\underline{\omega}$ with $|\underline{\omega}'|=|\underline{\omega}|=n$ if
$\omega_1 = \omega'_1,\dots,\omega_{n-1}=\omega'_{n-1}$ and 
$\omega_n>\omega'_n$.
A pair of random variables $V(\underline{\omega})$ and $h(\underline{\omega})$ 
is attached at each node.
All these variables are statistically independent and have marginal
distributions $p(h)$  (the $h(\underline{\omega})$'s) and 
$P(V)$  (the $V(\underline{\omega})$'s). A directed polymer 
is given by a pair of nodes $\underline{\omega}^1\prec \underline{\omega}^2$.
To each polymer we assign an energy as follows:
\begin{eqnarray}
E(\underline{\omega}^1, \underline{\omega}^2) =  
\sum_{\underline{\omega}^1\preceq \underline{\omega}\prec\underline{\omega}^2}
V(\underline{\omega})+
\sum_{\underline{\omega}^1\prec \underline{\omega}\preceq\underline{\omega}^2}
\;
\sum_{\underline{\omega}':\;
\underline{\omega}'\mbox{\tiny{ is an older}}\atop\mbox{\tiny{brother of }} 
\underline{\omega}}
h(\underline{\omega}')
\end{eqnarray}
Moreover we use the shorthand $E(\underline{0},\underline{\omega})\to
E(\underline{\omega})$ and define the following partition functions:
\begin{eqnarray}
Z_n(\beta)\equiv\sum_{\underline{\omega}:\;|\underline{\omega}|=n}
e^{-\beta E(\underline{\omega})}\\
Z_n(\beta|\underline{\omega}) = 
\sum_{\underline{\omega}'\succeq \underline{\omega}:\atop
|\underline{\omega}'|-|\underline{\omega}|=n}
e^{-\beta E(\underline{\omega},\underline{\omega}')}
\end{eqnarray}
The velocity of the front wave studied in the previous paragraphs corresponds in
this language to the random variable:
\begin{eqnarray}
c(\beta) \equiv \lim_{n\to\infty}\frac{1}{n\beta} \log Z_n(\beta)
\label{FreeEnergy}
\end{eqnarray}
The model has two phases. In the high temperature phase ($\beta\le\beta_c$)
the fluctuations of $Z_n(\beta)$ are small and 
\begin{eqnarray}
c(\beta) = \lim_{n\to\infty}\frac{1}{n\beta} \log \<Z_n(\beta)\> = v(\beta)
\label{HighTFreeEnergy}
\end{eqnarray}
In the low temperature phase ($\beta>\beta_c$) 
the fluctuations  become large and $c(\beta)$ is
fixed by simple convexity and monotonicity arguments. The key point 
of the approach used in Ref. \cite{BuffetTrees} is to estimate these 
fluctuations by proving that, for $\beta<\beta_c$:
\begin{eqnarray}
\frac{\<Z_n(\beta)^{\alpha}\>}{\<Z_n(\beta)\>^{\alpha}}\le 
\mbox{Bound}(\alpha,\beta)
\label{Bound}
\end{eqnarray}
for some $1<\alpha<2$ uniformly in $n$. This is enough for obtaining 
Eq. (\ref{HighTFreeEnergy}). 

Let us define the normalized variables 
$M_n(\beta)\equiv Z_n(\beta)/\<Z_n(\beta)\>$.
In Ref. \cite{BuffetTrees} the bound in Eq. (\ref{Bound}) is obtained
starting with the second moment of $M_n(\beta)$, and then refining the 
inequality for the fractional moments of order $1<\alpha<2$. Notice that 
looking at the $m$-th moment of the partition function is a well known method 
\cite{DerridaREM} for obtaining an upper estimate on the critical 
temperature (the estimate becomes worser as $m$ gets larger).
Let us have a look at the first two integer moments:
\begin{eqnarray}
\<Z_{n+1}(\beta)\> &= &\<e^{-\beta V}\>\sum_{k=0}^{\infty}\<e^{-\beta h}\>^k\;
\<Z_n(\beta)\>\\
\<Z_{n+1}^2(\beta)\> &= & \left(\<e^{-\beta V}\>
\sum_{k=0}^{\infty}\<e^{-\beta h}\>^k\right)^2\;
\left[\<Z_n^2(\beta)\>-Z_n(2\beta)\right]+\nonumber\\
&&+\<e^{-2\beta V}\>\sum_{k=0}^{\infty}\<e^{-2\beta h}\>^k
\left(1+2\sum_{l=1}^{\infty}\<e^{-\beta h}\>^l\right)\;
\<Z_n(2\beta)\>
\end{eqnarray}
In general the $m$-th moment is finite (but not necessarily uniformly bounded) 
only if $\<e^{-m\beta h}\><1$ i.e. if $\beta<\beta_1/m$. There is no 
integer moment of order greater than one which remains finite in the 
interval $(0,\beta_c)$. This fact forces us to a slight modification of the 
proof presented in Ref. \cite{BuffetTrees}.
We use the trivial identity:
\begin{eqnarray}
Z_{n+1}(\beta)=\sum_{\underline{\omega}:\;|\underline{\omega}|=1}
e^{-\beta E(\underline{\omega})}Z_n(\beta|\underline{\omega})
\end{eqnarray}
and estimate the $\alpha$-th moment (with $1<\alpha<2$) as follows:
\begin{eqnarray}
Z_{n+1}^{\alpha}(\beta) & = & 
\left\{
\sum_{\underline{\omega}^1:\atop |\underline{\omega}^1|=1} 
\sum_{\underline{\omega}^2:\atop |\underline{\omega}^2|=1}
e^{-\beta [E(\underline{\omega}^1)+E(\underline{\omega}^2)]}\;
Z_n(\beta|\underline{\omega}^1)\;Z_n(\beta|\underline{\omega}^2)
\right\}^{\alpha/2}\le\nonumber\\
&\le &\sum_{\underline{\omega}^1:\atop |\underline{\omega}^1|=1} 
\sum_{\underline{\omega}^2:\atop |\underline{\omega}^2|=1}
e^{-\frac{\alpha\beta}{2} [E(\underline{\omega}^1)+E(\underline{\omega}^2)]}\;
Z_n^{\alpha/2}(\beta|\underline{\omega}^1)\;
Z_n^{\alpha/2}(\beta|\underline{\omega}^2)
\end{eqnarray}
For a temperature such that $\alpha\beta<\beta_1$  we can take the averages 
and sum up the series:
\begin{eqnarray} 
\<Z_{n+1}^{\alpha}(\beta)\>&\le & 
\sum_{\underline{\omega}:\atop |\underline{\omega}|=1}
\<e^{-\alpha\beta E(\underline{\omega})}\>\;
\<Z_n^{\alpha}(\beta)\>+
\sum_{\underline{\omega}^1\neq\underline{\omega}^2:\atop 
|\underline{\omega}^i|=1}
\<e^{-\frac{\alpha\beta}{2} 
[E(\underline{\omega}^1)+E(\underline{\omega}^2)]}\>\;
\<Z_n^{\alpha/2}(\beta)\>^2\le\nonumber\\
&\le & \phi(\alpha\beta)\<Z_n^{\alpha}(\beta)\>+
2 \phi(\alpha\beta)\sum_{l=1}^{\infty}\<e^{-\frac{\alpha\beta}{2}h}\>^l
\<Z_n(\beta)\>^{\alpha}
\end{eqnarray}
Rewriting this formula for the normalized variables we get
\begin{eqnarray}
\< M_{n+1}^{\alpha}(\beta)\>&\le& 
\left[\frac{\phi(\alpha\beta)}{\phi(\beta)^{\alpha}}\right]
\< M_n^{\alpha}(\beta)\>+
2\left[\frac{\phi(\alpha\beta)}{\phi(\beta)^{\alpha}}\right]
\sum_{l=1}^{\infty}\<e^{-\frac{\alpha\beta}{2}h}\>\equiv\nonumber\\
&\equiv &\psi(\alpha,\beta)\< M_n^{\alpha}(\beta)\>
+\chi(\alpha,\beta)
\end{eqnarray}
At this point we observe, following Ref. \cite{BuffetTrees}, that, if
$\frac{dv}{d\beta}(\beta)<0$ (i.e. $\beta<\beta_c$) then we can choose 
$\alpha >1$ such that $\psi(\alpha,\beta)<1$. The condition to be imposed 
on $\alpha$ for obtaining this inequality is $\alpha<\beta_c/\beta$
(notice that this inequality implies the previous one $\alpha<\beta_1/\beta$).
The desired bound is obtained by using Gronwall lemma together 
with the fact that $\< M_0^{\alpha}(\beta)\>=1$:
\begin{eqnarray}
\< M_n^{\alpha}(\beta)\>\le \psi^n(\alpha,\beta)+
\frac{1-\psi^n(\alpha,\beta)}{1-\psi(\alpha,\beta)}\chi(\alpha,\beta)\le
1+\frac{1}{1-\psi(\alpha,\beta)}\chi(\alpha,\beta)
\end{eqnarray}

\clearpage

\end{document}